\documentclass{article}

\usepackage[preprint,nonatbib]{neurips_2022}

\usepackage[utf8]{inputenc} % allow utf-8 input
\usepackage[T1]{fontenc}    % use 8-bit T1 fonts
\usepackage{hyperref}       % hyperlinks
\usepackage{url}            % simple URL typesetting
\usepackage{booktabs}       % professional-quality tables
\usepackage{amsfonts}       % blackboard math symbols
\usepackage{nicefrac}       % compact symbols for 1/2, etc.
\usepackage{microtype}      % microtypography
\usepackage{xcolor}         % colors
\usepackage{tipa}           % English pronunciation
\usepackage{CJKutf8}        % Chinese font

\usepackage{graphicx}
\usepackage{float}
\usepackage{subfig}
\usepackage{overpic}
\usepackage{wrapfig}
\usepackage{caption}

\usepackage{multirow}
\usepackage{array,makecell,booktabs}
\newcolumntype{M}[1]{>{\centering\arraybackslash}m{#1}}
\usepackage{wrapfig}

\usepackage{appendix}

\usepackage{verbatim}

\title{Dict-TTS: Learning to Pronounce with Prior \\
Dictionary Knowledge for Text-to-Speech}

\author{%
  Ziyue Jiang\thanks{Equal contribution.} \\
  Zhejiang University\\
  \texttt{ziyuejiang@zju.edu.cn} \\
   \And
   Zhe Su\footnotemark[1] \\
   Zhejiang University \\
   \texttt{suzhesz00@gmail.com} \\
   \And
   Zhou Zhao\thanks{Corresponding author} \\
   Zhejiang University \\
   \texttt{zhaozhou@zju.edu.cn} \\
   \AND
   Qian Yang \\
   Zhejiang University \\
   \texttt{qyang1021@foxmail.com} \\
   \And
   Yi Ren \\
   Bytedance AI Lab \\
   \texttt{ren.yi@bytedance.com} \\
   \And
   Jinglin Liu \\
   Zhejiang University \\
   \texttt{jinglinliu@zju.edu.cn} \\   
   \AND
   Zhenhui Ye \\
   Zhejiang University \\
   \texttt{zhenhuiye@zju.edu.cn} \\   
}

\begin{document}

\maketitle

\begin{abstract}
Polyphone disambiguation aims to capture accurate pronunciation knowledge from natural text sequences for reliable Text-to-speech (TTS) systems. However, previous approaches require substantial annotated training data and additional efforts from language experts, making it difficult to extend high-quality neural TTS systems to out-of-domain daily conversations and countless languages worldwide. This paper tackles the polyphone disambiguation problem from a concise and novel perspective: we propose Dict-TTS, a semantic-aware generative text-to-speech model with an online website dictionary (the existing prior information in the natural language). Specifically, we design a semantics-to-pronunciation attention (S2PA) module to match the semantic patterns between the input text sequence and the prior semantics in the dictionary and obtain the corresponding pronunciations; The S2PA module can be easily trained with the end-to-end TTS model without any annotated phoneme labels. Experimental results in three languages show that our model outperforms several strong baseline models in terms of pronunciation accuracy and improves the prosody modeling of TTS systems. Further extensive analyses demonstrate that each design in Dict-TTS is effective. The code is available at \url{https://github.com/Zain-Jiang/Dict-TTS}.
 
%Local changes in pitch and speaking duration can convey semantic meaning, while global properties such as overall pitch trajectory can convey mood and emotion.

%Pre-trained word embeddings from massive unlabeled corpora offer a compact way of injecting a prior notion of word similarity into models that would otherwise treat words as discrete, isolated categories. However, the specific properties of language captured by any particular embedding scheme can be difficult to control, and, further, may not be ideally suited to the task at hand. 《Unsupervised Learning of Syntactic Structure with Invertible Neural Projections》

\end{abstract}

%In recent years, deep-learning-based text-to-speech (TTS) models have succeeded in synthesizing high-quality speech from the given text and have attracted a lot of attention in speech community~\cite{arik2017deep,li2019neural,ren2019fastspeech,shen2018natural,wang2017tacotron,kim2020glow,DBLP:conf/interspeech/EliasZS0JSW21,ren2021portaspeech}. 
\section{Introduction}
Capturing the pronunciations from raw texts is challenging for end-to-end text-to-speech (TTS) systems~\cite{arik2017deep,li2019neural,ren2019fastspeech,shen2018natural,wang2017tacotron,kim2020glow,DBLP:conf/interspeech/EliasZS0JSW21,ren2021portaspeech,huang2022prodiff,huang2022generspeech}, since there are full of words that are not covered by general pronunciation rules~\cite{black1998issues,jensen2004principles,tan2021survey}. Therefore, polyphone\footnote{Polyphones are characters having more than one phonetic value. See Appendix \ref{definition_hp} for further details.} disambiguation (one of the biggest challenges in converting texts into phonemes~\cite{g2pE2019,yu2020multilingual,sokolov2020neural}) plays an important role in the construction of high-quality neural TTS systems~\cite{hida2022polyphone,park2020g2pm}. However, since the exact pronunciation of a polyphone must be inferred based on its semantic contexts, current solutions still face several challenges: 1) the rule-based approaches~\cite{zhang2001disambiguation,huang2008disambiguating} with limited linguistic knowledge or the neural models~\cite{yu2020multilingual,park2020g2pm,sokolov2020neural} trained on limited data suffer from significant performance degradation on out-of-domain text datasets; 2) the neural network-based models~\cite{chae2018convolutional,DBLP:conf/interspeech/CaiYZQL19,park2020g2pm} learn the grapheme-to-phoneme (G2P) mapping in an end-to-end manner without explicit semantics modeling, which hinders their pronunciation accuracy in real-life applications. 3) based on the above two points, a reliable polyphone disambiguation module is usually based on a combination of hand-crafted rules, structured G2P-oriented lexicons, and neural models~\cite{he2021neural}, which requires substantial phonemes labels and external knowledge from language experts.

\begin{figure}[tbp]
  \centering
    \includegraphics[scale=0.44]{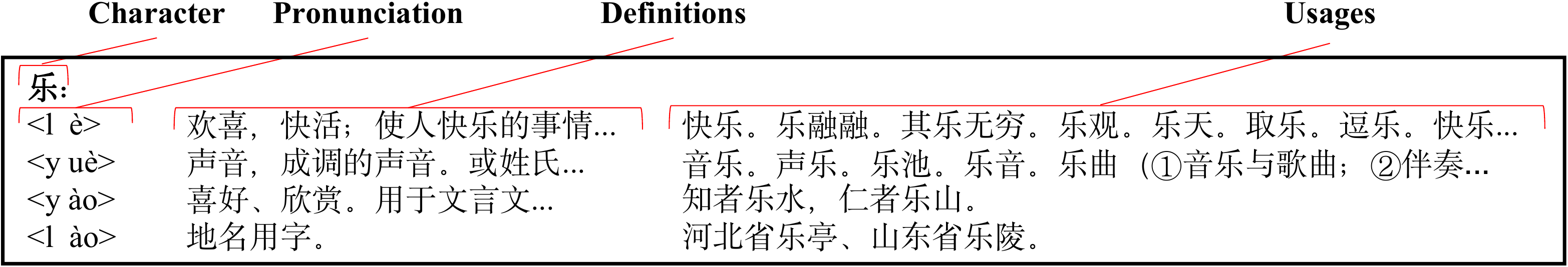}
  \caption{The illustration of the dictionary entry that contains information on character's or word's definitions, usages, and pronunciations. For example, in the Chinese sentence \begin{CJK}{UTF8}{gbsn}``快乐的人们''\end{CJK}, the pronunciation of the polyphone \begin{CJK}{UTF8}{gbsn}``乐''\end{CJK} should be inferred based on its semantic contexts.}
  \label{dictionary}
\end{figure}
Unlike the previous rule-based or neural network-based approaches, we address the above challenges with the existing prior information worldwide. As shown in Figure~\ref{dictionary}, an arbitrary dictionary used in daily life can be viewed as a prior knowledge database. Intuitively, it contains valuable prior knowledge for pronunciations in conversations. When one is confused about the acoustic pronunciation of a specific polyphone, he or she will resort to the dictionary website to infer its exact reading based on the semantic context. We imitate this scenario in our architecture design and propose Dict-TTS, an unsupervised polyphone disambiguation framework, which explicitly consults the online dictionary to identify the correct semantic meanings and acoustic pronunciations of polyphones. Specifically,

\begin{itemize}

\item To explicitly learn the semantics-to-pronunciation mapping, we adopt a semantic encoder to obtain the semantic contexts of the input text and utilize a semantics-to-pronunciation attention (S2PA) module to search the matched semantic patterns in the dictionary so as to find the correct pronunciations. We also use the retrieved semantic information as auxiliary information for prosody modeling of the TTS model.

%Then the semantic representations and the embeddings of the selected pronunciations are fed into a linguistic encoder for feature fusion.

\item To perform polyphone disambiguation without phoneme labels, we combine our S2PA module into end-to-end TTS systems' training and inference processes. Different from current neural polyphone disambiguation models, our module can be trained with the guidance of mel-spectrogram reconstruction loss in a fully end-to-end manner, which significantly reduces the cost of building such a system.

\end{itemize}
To demonstrate the generalization ability of our Dict-TTS, we perform experiments on three datasets, including a standard Mandarin dataset~\cite{baker2017chinese}, a Japanese corpus~\cite{sonobe2017jsut}, and a Cantonese dataset~\cite{ardila2020common}. Experiments on these datasets show that Dict-TTS outperforms other state-of-the-art polyphone disambiguation models in pronunciation accuracy and improves the prosody modeling of TTS systems in terms of both subjective and objective evaluation metrics. The pronunciation accuracy of Dict-TTS is further improved by being pre-trained on a large-scaled automatic speech recognition (ASR) dataset. The main contributions of this work are summarized as follows:

\begin{itemize}

\item We incorporate the online dictionary into TTS systems and propose a semantic-aware method for polyphone disambiguation, which improves the pronunciation accuracy and robustness of end-to-end TTS systems. Moreover, the idea of introducing the prior knowledge worldwide can also inspire other tasks like neural language modeling~\cite{khandelwal2019generalization} and sequence labeling~\cite{liu2021lexicon}.

\item We propose a novel and general framework for unsupervised polyphone disambiguation in TTS systems, which further enables the efficient pre-training on large-scaled ASR datasets and improves the generalization capacity significantly. 

\item We also find that the retrieved semantics in the dictionary knowledge can be used as auxiliary information to improve prosody modeling and help the TTS system to generate more expressive speech.

\item We further analyze the characteristics of the linguistic encoder based on phoneme and character and provide valuable interpretations about our semantics-to-pronunciation attention module.

\end{itemize}

%``read'' is pronounced /ri:d/ and /red/ in the present and past tenses, respectively.

%However, collecting a complete speech dataset that covers all necessary knowledge is extremely costly and it is also challenging for TTS models to learn the knowledge.

%a large amount of data is generally required to cover all knowledge necessary to produce speech: pronunciation of each grapheme under different contexts, semantic meanings of various sentences to predict natural prosody, etc.

\section{Background}
This section describes the background of TTS, grapheme-to-phoneme (G2P) pipeline, and their relations with polyphone disambiguation. We also review the existing works that aim at semantic-aware polyphone disambiguation and analyze their advantages and disadvantages.

\paragraph{Text-to-Speech}
Text-to-speech (TTS) models~\cite{wang2017tacotron,arik2017deep,lee2018learning,li2019neural,ren2019fastspeech,kim2020glow,ren2021portaspeech,liu2022diffsinger} first generate mel-spectrogram from text and then synthesize speech waveform from the generated mel-spectrogram using a separately pre-trained vocoder~\cite{DBLP:conf/ssw/OordDZSVGKSK16,kong2020hifi,huang2022fastdiff}, or directly generate waveform from text in an end-to-end manner~\cite{DBLP:conf/iclr/0006H0QZZL21,DBLP:conf/iclr/DonahueDBES21,kim2021conditional}. The frontend model of end-to-end TTS system should tackle one important task, i.e., polyphone disambiguation~\cite{zhang2020unified}. Properly mapping the grapheme sequence into phoneme sequence requires the linguistic encoder to capture the empirical pronunciation rules in daily conversations. However, it is extremely difficult for the linguistic encoder to learn all the pronunciation rules necessary to produce speech in an end-to-end manner, which results in inevitable mispronunciations in the generated speech. To alleviate this problem, robust TTS models usually convert the text sequence into the phoneme sequence with open-source grapheme-to-phoneme pipelines and predict the mel-spectrogram from the phoneme sequence. However, the rule-based or neural network-based grapheme-to-phoneme pipelines suffer from significant performance degradation on out-of-domain datasets since it is extremely difficult and costly for them to cover all linguistic knowledge.

\paragraph{Grapheme-to-Phoneme} The grapheme-to-phoneme models map grapheme sequence to phoneme sequence to reduce pronunciation errors in modern TTS systems. For logographic languages like Chinese, Japanese and Korean, although the lexicon can cover nearly all the characters, there are full of polyphones that can only be decided according to the semantic context of a character~\cite{tan2021survey}. Thus, polyphone disambiguation is the most important challenge in grapheme-to-phoneme conversions for this kind of languages. Moreover, many alphabetic languages including English and French also have polyphones in daily conversations. Current polyphone disambiguation approaches can be categorized into the rule-based approach~\cite{zhang2001disambiguation,huang2008disambiguating} and the data-driven approach~\cite{shan2016bi,chae2018convolutional,DBLP:conf/interspeech/CaiYZQL19,park2020g2pm}. The rule-based G2P method is based on a combination of hand-crafted rules and structured G2P-oriented lexicons, which requires a substantial amount of linguistic knowledge. The data-driven G2P algorithm adopts statistical methods~\cite{liu2010polyphonic} or neural encoder-decoder architecture~\cite{DBLP:conf/interspeech/YaoZ15,DBLP:conf/interspeech/CaiYZQL19,sun2019token,yu2020multilingual,park2020g2pm,sokolov2020neural,wang2021joint}. However, building a data-driven G2P model requires a large amount of carefully labeled data and substantial linguistic knowledge from language experts, which is extremely costly and laborious.

\paragraph{Semantic-Aware Polyphone Disambiguation} Polyphone disambiguation is the core issue for G2P conversion in various languages. The pronunciation of a polyphone is defined by the semantic context information of neighbouring characters~\cite{tan2021survey}. In order to comprehend the semantic meaning in the given sentence for polyphone disambiguation, previous methods~\cite{dai2019disambiguation,yang2019pre,sun2019knowledge,hida2022polyphone,chen2022g2pw} have adopted the pre-trained language model~\cite{devlin2018bert} to extract semantic features from raw character sequences and predict the pronunciation of polyphones with neural classifiers according to the semantic features. Among them, PnG BERT and Mixed-Phoneme BERT~\cite{DBLP:conf/interspeech/JiaZSZW21,zhang2022mixed} take both phoneme and grapheme as input to train an augmented BERT and use the pre-trained augmented BERT as the TTS encoder. However, these methods still require annotated data to train and can not be incorporated into the TTS training in an end-to-end manner. Although a concurrent work, NLR~\cite{he2021neural}, directly injects BERT-derived knowledge into the TTS systems without phoneme labels and successfully reduces pronunciation errors, their method confounds the acoustic and semantic space, which significantly affects the pronunciation accuracy\footnote{The detailed differences between NLR~\cite{he2021neural} and our method can be found in Appendix~\ref{diff_nlr}}.

\section{Method}
To exploit the prior linguistic knowledge in the online dictionary for TTS systems, we propose Dict-TTS, which explicitly captures the semantic relevance between the input sentences and the dictionary entries for polyphone disambiguation. In this section, we firstly introduce the overall architecture design of Dict-TTS based on PortaSpeech~\cite{ren2021portaspeech}. Then after the comparison between the phoneme-based and character-based TTS systems in the acoustic and semantic space, we design a novel semantics-to-pronunciation attention (S2PA) module to learn grapheme-to-phoneme mappings based on the semantic context; In general, Dict-TTS exploits the prior pronunciation knowledge in the online dictionary with the following steps: Firstly, the character sequence is fed into the self-attention based semantic encoder to obtain the semantic representations of the input character sequence, and we utilize a pre-trained cross-lingual language model~\cite{conneau2020unsupervised} to extract the semantic context information in the dictionary entries; Secondly, we calculate the most relevant dictionary entries of the input graphemes and obtain the corresponding pronunciation sequence contained in the dictionary entries; Finally, the extracted semantic context information and the pronunciations are fed into the linguistic encoder for feature fusion. We describe these designs in detail in the following subsections.

\begin{figure}[tbp]
  \centering
    \includegraphics[scale=0.44]{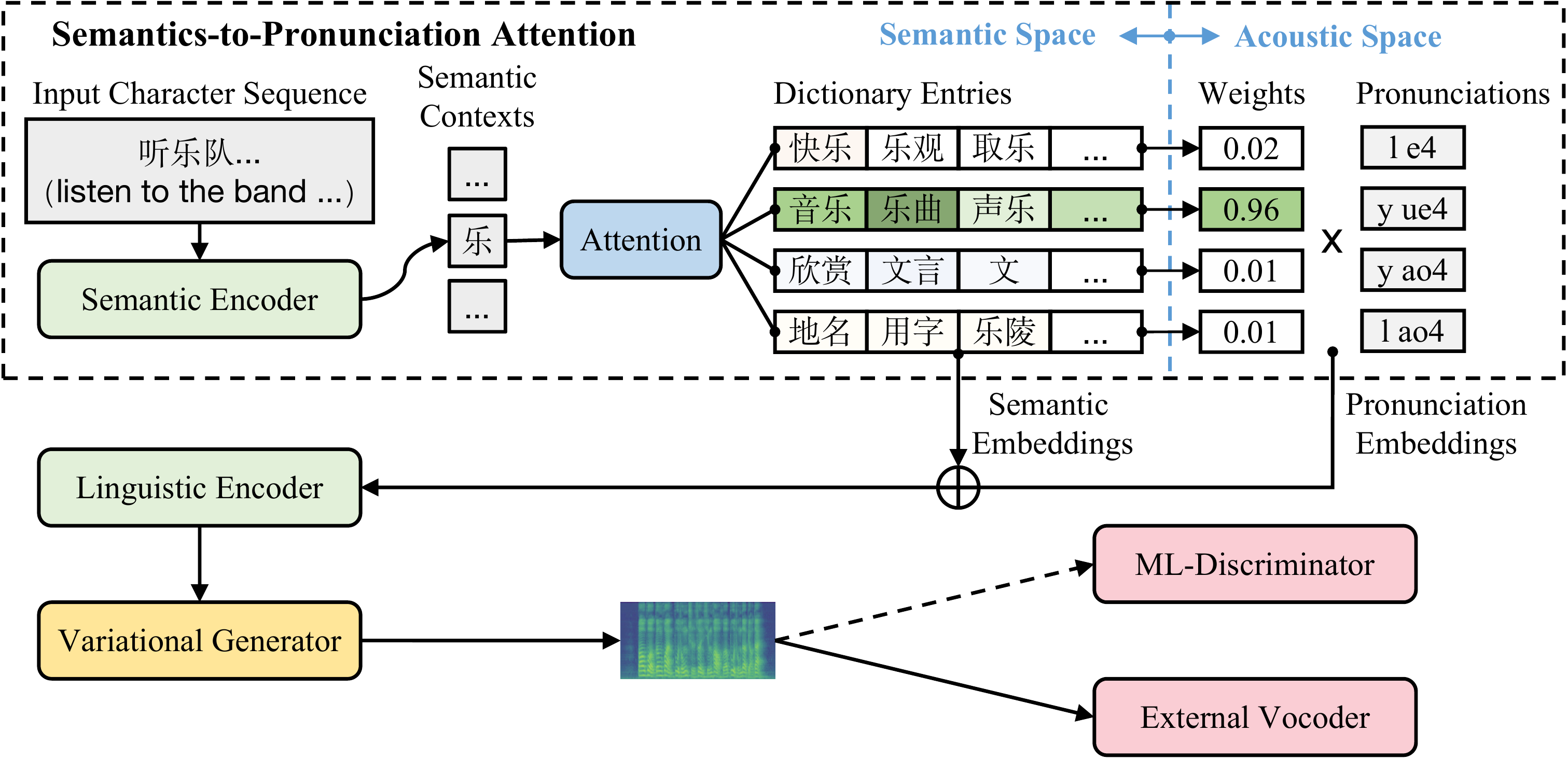}
  \caption{The overall architecture for Dict-TTS. The character \begin{CJK}{UTF8}{gbsn}``乐''\end{CJK} has 4 possible pronunciations coupled with 4 different meanings. The module inside the dashed box is our semantics-to-pronunciation attention (S2PA). The S2PA module measures the semantic similarity between the input character and the corresponding dictionary entries and aggregates the attention weights into pronunciation weights. Then the weighted semantic embeddings and pronunciation embeddings are fed into the linguistic encoder for feature fusion. The semantic and acoustic spaces described in Subsection~\ref{comparsion_phoneme_character} are separated by the blue dashed line. ``ML-Discriminator'' denotes Multi-Length Discriminator in HiFiSinger~\cite{DBLP:journals/corr/abs-2009-01776}. The dashed black line denotes that the operation is only executed in the training phase.}
  \label{main_architecture_1}
\end{figure}
\subsection{Model Overview}
The overall model architecture of Dict-TTS is shown in Figure \ref{main_architecture_1}. Dict-TTS keeps the main structures of PortaSpeech: a Transformer-based linguistic encoder; a VAE-based variational generator with flow-based prior to generate diverse mel-spectrogram. The flow-based post-net in PortaSpeech is replaced with a multi-length discriminator~\cite{DBLP:journals/corr/abs-2009-01776} based on random windows of different lengths, which has been proved to improve the naturalness of word pronunciations~\cite{2022arXiv220411792Y}. However, since there are no phoneme inputs in our scenarios, we replace the linguistic encoder that combines hard word-level alignment and soft phoneme-level alignment with: 1) a semantic encoder that extracts the semantic representations in the grapheme sequence; 2) a semantics-to-pronunciation attention module that matches the semantic patterns between the dictionary entries and the grapheme representations and obtains the corresponding semantic embedding and pronunciation embedding; 3) a linguistic encoder that fuses the semantic embedding and pronunciation embedding.

\subsection{Comparison between the phoneme-based and character-based TTS systems}
\label{comparsion_phoneme_character}
In this subsection, we make preliminary analyses about the linguistic encoder of phoneme-based and character-based TTS systems. For simplicity, we describe the following concepts according to the logographic writing system, where a single written character represents a complete grammatical word or morpheme. These concepts can be extended to alphabetic languages like English by replacing ``character'' with ``word''.

\begin{wrapfigure}[14]{r}{14em}
  \centering
    \includegraphics[scale=0.45]{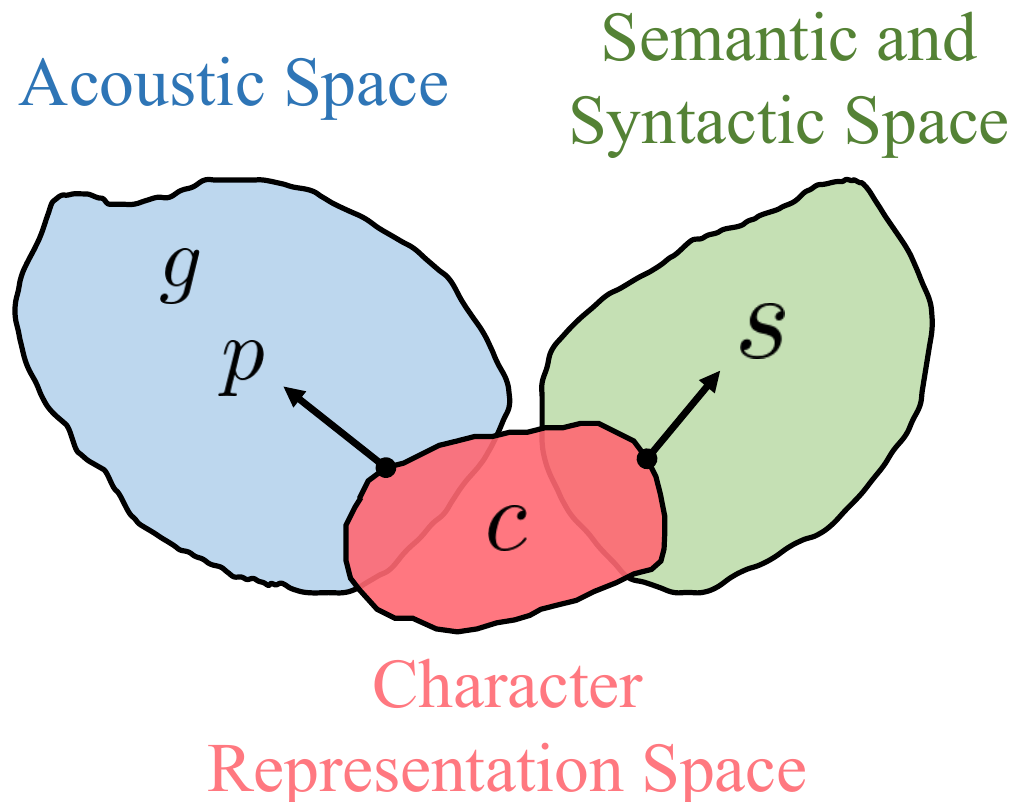}
  \caption{The illustration of representations in the linguistic encoder.}
  \label{two_space}
\end{wrapfigure}
\paragraph{Phoneme-based TTS systems} As is shown in Figure \ref{two_space}, the linguistic encoder of the phoneme-based TTS system takes phoneme sequence $p$ generated by the G2P module as inputs. The main challenge of the phoneme-based linguistic encoder is to comprehend the semantic and syntactic representation $s$ from $p$ and deduce the pitch trajectory, speaking duration, and other acoustic features from $s$ and $p$ to generate the expressive and natural pronunciation hidden state $g$. Since the phoneme sequence $p$ is a combination of the smallest units of sound in speech, it may be ambiguous in terms of semantic meaning, which brings difficulties for the deduction of the representation $s$. For example, ``AE1, T, F, ER1, S, T'' can be easily classified as ``At first'', but ``W, EH1, DH, ER0'' can be classified as ``Whether'' or ``Weather''. Homophones like ``to'', ``too'', and ``two'' can be converted to the same phoneme sequence ``T, UW1'', but their local speaking duration and pitch are different. Moreover, the tree-structured syntactic information contained in the word-based input sequence is missing. Since semantic information and syntactic information possess rich intonational features such as pitch accent and phrasing of the input text~\cite{2022arXiv220411792Y}, the ambiguity of $s$ hurts the prosody modeling in phoneme-based TTS systems.

\paragraph{Character-based TTS systems} 
The first challenge for a character-based TTS system is to predict the correct phoneme sequence $p$. Unlike the phoneme-based TTS system, the character-based TTS system does not know the phoneme sequence $p$ when characters arrive. Thus, it does not know how to pronounce the words accurately at first. Then the mel-spectrogram reconstruction loss in TTS training would drag the character representation $c$ to the acoustic space. For example, Chinese characters \begin{CJK}{UTF8}{gbsn}``火''\end{CJK} (``fire'' in English) and \begin{CJK}{UTF8}{gbsn}``伙''\end{CJK} (``partner'' in English) share the same pronunciation (``H UO3'') and have different semantic meanings. However, with the guidance of the mel-spectrogram reconstruction loss, their representations distribute according to the acoustic pronunciation, which hinders the semantic comprehension for polyphone disambiguation and prosody modeling. 

From the above analyses, it can be seen that the character representations $c$ should locate in the semantic space so that we can easily capture $s$ based on the context, deduce $p$ based on the dictionary and $s$, and finally obtain the natural pronunciation hidden state $g$. The following subsection mainly describes how we achieve the above goals with our semantics-to-pronunciation attention module.

\subsection{Semantics-to-Pronunciation Attention}
\label{method_S2PA}
\begin{wrapfigure}[13]{r}{12em}
  \centering
    \includegraphics[scale=0.32]{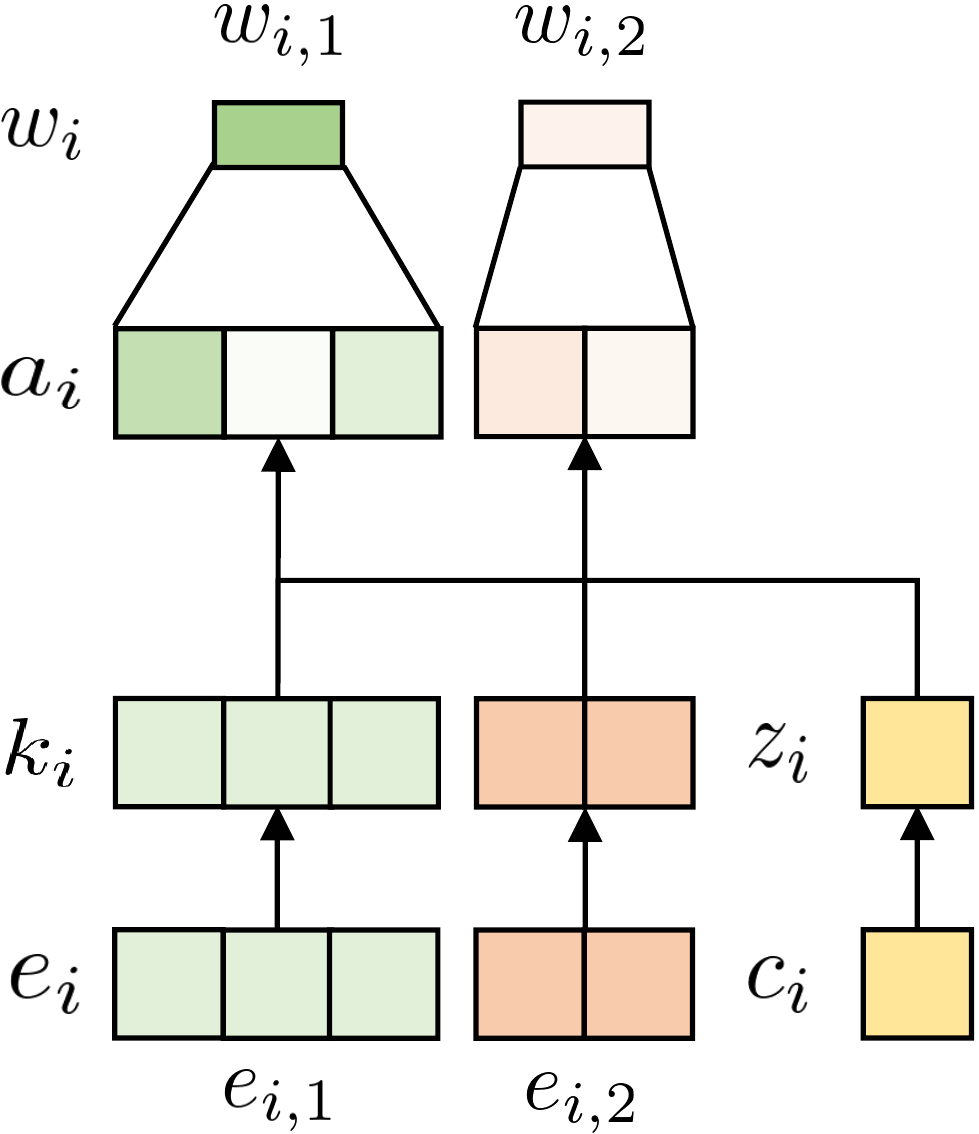}
  \caption{The illustration of semantic pattern matching.}
  \label{S2PA}
\end{wrapfigure}
As shown in Figure \ref{main_architecture_1}, the semantics-to-pronunciation attention (S2PA) module is designed for explicit semantics comprehension and polyphone disambiguation. 

\paragraph{Dictionary Definition} Assuming that we have an dictionary $D$ which contains a sequence of characters $C=[c_1, c_2, ..., c_n]$, where $n$ is the size of characters set in a language. The dictionary can be easily downloaded from online websites (see Appendix~\ref{dictionary_detail} for further details). In the dictionary, each character $c_{i}$ has a sequence of possible pronunciations $p_{i}=[p_{i,1},p_{i,2},...,p_{i,m}]$ and each pronunciation $p_{i,j}$ has its corresponding dictionary entry $e_{i,j}=[e_{i,j,1},e_{i,j,2},...,e_{i,j,u}] \in E$ (definitions, usages, and translations are merged together as a single characters sequence), where $m$ is the number of the possible pronunciation of $c_{i}$ and $u$ is the number of characters in the corresponding entry. Note that for polyphones $m>1$ and for characters that have only one pronunciation $m=1$. 

\paragraph{Semantics Matching} The goal of our S2PA is to obtain the pronunciation sequence $p$ by measuring the semantic similarity between the input character sequence $t=[t_{1},...,t_{l}]$ and the corresponding gloss items in the dictionary, where $l$ is the sequence length. As shown in Figure~\ref{S2PA}, we firstly extract the semantic context information $\mathbf{k}$ of every entry $e$ with a pretrained cross-lingual language model~\cite{conneau2020unsupervised} and store $\mathbf{k}$ as the prior dictionary knowledge. Then we use a semantic encoder to obtain the semantic contexts $\mathbf{z}$ from the input character sequence $t$. For each character token $t_{i}$, its semantic feature vector $\mathbf{z}_{i}$ is used as the query vector to a semantics-based attention module. Here, attention is used to learn a similarity measure between the semantic feature vector $\mathbf{z}_{i}$ and $\mathbf{k}_{i,j,k}$:
\begin{equation}
   [\mathbf{a}_{i,1,1}, ..., \mathbf{a}_{i,m,u}] = \frac{[\mathbf{k}_{i,1,1}, ..., \mathbf{k}_{i,m,u}] \cdot \mathbf{z}_{i}^\top}{d} \ , 
\end{equation}
where $d$ is the scaling factor and $[\mathbf{a}_{i,1,1},...,\mathbf{a}_{i,m,u}]$ denotes the semantic similarity between $t_{i}$ and each item in $[e_{i,1,1},e_{i,m,u}]$. The retrieved semantic embeddings $\mathbf{s^{\prime}}_{i}$ can be extracted by $\mathbf{s^{\prime}}_{i}=softmax([\mathbf{a}_{i,1,1}, ..., \mathbf{a}_{i,m,u}]) \cdot [\mathbf{k}_{i,1,1}, ..., \mathbf{k}_{i,m,u}]$. The rich linguistic information in $\mathbf{s^{\prime}}_{i}$ can be used as auxiliary information to improve the naturalness and expressiveness of the generated speeches. 

\paragraph{Polyphone Disambiguation} The aggregated attention weight $\mathbf{w}_{i,j} = \sum_{k=1}^u\mathbf{a}_{i,j,k}$ can be seen as the probability of the pronunciation $p_{i,j}$. However, since a polyphone in a specific sentence has only one correct pronunciation, here we use the Gumbel-Softmax function~\cite{jang2016categorical} to sample a differentiable approximation of the most possible pronunciation $p_{i}^{\prime}$:
\begin{equation}
    w_{i,j}
    =\frac{\exp \left(\left(\log \left(\mathbf{w}_{i,j}\right)+g_{i,j}\right) / \tau\right)}{\sum_{l=1}^{m} \exp \left(\left(\log \left(\mathbf{w}_{i,l}\right)+g_{i,l}\right) / \tau\right)} \ , 
\end{equation}
\begin{equation}
    p_{i}^{\prime} = \sum_{j=1}^{m} w_{i,j} \cdot p_{i,j} \ , 
\end{equation}
where $g_{i,1}, ..., _{i,m}$ are i.i.d samples drawn from Gumbel(0,1) distribution and $\tau$ is the softmax temperature. Then the retrieved pronunciation embeddings $p_{i}^{\prime}$ and semantic embeddings $\mathbf{s^{\prime}}_{i}$ are then fed into the rest of the linguistic encoder for feature fusion and syntax prediction. Intuitively, our S2PA module can be thought of as an end-to-end method for decomposing the character representation, pronunciation, and semantics with dictionary following the preliminary analyses in Subsection~\ref{comparsion_phoneme_character}. Through our S2PA module, the character representation is successfully distributed in the semantic space so that the model can easily deduce the correct pronunciation and semantics based on the lexicon knowledge like human brain. 

\subsection{Training and Pre-training}
In training, the S2PA module weights (including character token embeddings and the pronunciation token embeddings) are jointly trained by the reconstruction loss from the TTS decoder. Thus, our Dict-TTS does not require any explicit phoneme labels. In inference, the pronunciations can be specified by feeding the text sequence into the S2PA module to get the predicted pronunciations. Besides, our method is compatible with the predefined rules from language experts by directly adding specific rules to pronunciation weight $w_{i,j}$. 

Although the semantics-to-pronunciation mappings can be explicitly learned by the S2PA module, it could be not accurate enough, due to the following reason: the text training data is not large enough (about 10k sentences) in the TTS dataset, leading to relatively inaccurate context comprehension. To improve the semantic comprehension and the generalization capacity for our S2PA module, we propose a pre-training method using low-quality text-speech pairs from large-scaled automatic speech recognition (ASR) datasets. Since our S2PA module can be trained without hand-crafted phoneme labels, it can be easily pre-trained and effectively finetuned to various domains to improve the pronunciation accuracy of the TTS systems.

\section{Experiments}
\subsection{Experimental Setup}
\label{Experimental Setup}
\paragraph{Datasets} We evaluate Dict-TTS on three
datasets of different sizes, including: 1) Biaobei~\cite{baker2017chinese}, a Chinese speech corpus consisting of 10,000 sentences (about 12
hours) from a Chinese speaker; 2) JSUT~\cite{sonobe2017jsut}, a Japanese speech corpus containing reading-style speeches from a Japanese female speaker (we use the basic5000 subset that contains 5,000 daily-use sentences). 3) Common Voice (HK)~\cite{ardila2020common}, a Cantonese speech corpus that contains 125 hours of speeches from 2,869 speakers (We use the 102 hours of the validated speeches). For each of the three datasets, we randomly sample 400 samples for validation and 400 samples for testing. We randomly choose 50 samples in the test set for subjective audio quality and prosody evaluation and use all testing samples for other evaluations. The ground truth mel-spectrograms are generated from the raw waveform with the frame size 1024 and the hop size 256. For computational efficiency, we firstly use the pre-trained XLM-R~\cite{conneau2020unsupervised} model to extract the semantic representations from the raw text of the whole dictionary and record them in the disk. We then load the mini-batch along with the pre-constructed dictionary representation during training and testing. 

\paragraph{Implementation Details} 
\label{Implementation Details}
Our Dict-TTS consists of a S2PA module, an encoder, a variational generator and a post-net. The encoder consists of multiple feed-forward Transformer blocks~\cite{ren2019fastspeech} with relative position
encoding~\cite{shaw2018self} following Glow-TTS~\cite{kim2020glow}. The encoder and decoder in variational generator are 2D-convolution networks following PortaSpeech~\cite{ren2021portaspeech}. We replace the post-net in PortaSpeech with a multi-length discriminator~\cite{DBLP:journals/corr/abs-2009-01776}, which has been proved to improve the naturalness of word pronunciations~\cite{2022arXiv220411792Y}. We add more detailed model configurations in Appendix~\ref{dicttts_config},~\ref{baseline_config}. We train Dict-TTS on 1 NVIDIA 3080Ti GPU, with batch size of 40 sentences on each GPU. We use the Adam optimizer with $\beta_1 = 0.9$, $\beta_2 = 0.98$, $\epsilon = 10^{-9}$ and follow the same learning rate schedule in~\cite{vaswani2017attention}. The softmax temperature $\tau$ is initialized and annealed using the schedule in ~\cite{jang2016categorical}. It takes 320k steps for training until convergence. The predicted mel-spectrograms are transformed into audio samples using pre-trained HiFi-GAN~\cite{kong2020hifi}\footnote{\url{https://github.com/jik876/hifi-gan}}.

\begin{table}[]\small
\caption{The objective and subjective pronunciation accuracy comparisons. PER-O denotes phoneme error rate in the objective evaluation, PER-S denotes phoneme error rate in the subjective evaluation and SER-S denotes sentence error rate in the subjective evaluation.}
\label{table_1}
\centering
\begin{tabular}{@{}l|ccc|ccc|ccc@{}}
\toprule
\multirow{2}{*}{Method} & \multicolumn{3}{c|}{Biaobei}                                    & \multicolumn{3}{c|}{JSUT}                                       & \multicolumn{3}{c}{Common Voice (HK)}                           \\ \cmidrule(l){2-10} 
                        & \multicolumn{1}{c|}{PER-O} & \multicolumn{1}{c|}{PER-S} & SER-S & \multicolumn{1}{c|}{PER-O} & \multicolumn{1}{c|}{PER-S} & SER-S & \multicolumn{1}{c|}{PER-O} & \multicolumn{1}{c|}{PER-S} & SER-S \\ \midrule
Character                    & \multicolumn{1}{c|}{-}     & \multicolumn{1}{c|}{3.73\%}  & 30.50\%  & \multicolumn{1}{c|}{-}     & \multicolumn{1}{c|}{13.78\%}      &   65.50\%    & \multicolumn{1}{c|}{-}     & \multicolumn{1}{c|}{1.89\%} & 15.50\% \\
Phoneme                 & \multicolumn{1}{c|}{2.78\%}  & \multicolumn{1}{c|}{1.14\%}  & 7.00\%   & \multicolumn{1}{c|}{\textbf{1.55\%}}      & \multicolumn{1}{c|}{\textbf{0.92\%}}      &   \textbf{4.25\%}    & \multicolumn{1}{c|}{-}      & \multicolumn{1}{c|}{1.45\%}      &   10.25\%    \\
Dict-TTS                & \multicolumn{1}{c|}{\textbf{2.12\%}}  & \multicolumn{1}{c|}{\textbf{1.08\%}}  & \textbf{6.50\%}   & \multicolumn{1}{c|}{3.73\%}      & \multicolumn{1}{c|}{2.57\%}      & 22.75\%      & \multicolumn{1}{c|}{-}      & \multicolumn{1}{c|}{\textbf{1.23\%}} & \textbf{9.75\%} \\ \bottomrule
\end{tabular}
\end{table}

\subsection{Results of Pronunciation Accuracy}
\label{evaluation_pronunciation}
We compare the pronunciation accuracy of our Dict-TTS with other systems, including 1) character-based systems, where we directly feed character into the linguistic encoder; 2) Phoneme-based systems, where we convert the text sequence to the phoneme sequence~\cite{wang2017tacotron,shen2018natural,kim2020glow} with most popular open-source grapheme-to-phoneme tools\footnote{We use \textit{pypinyin} in the Biaobei dataset, \textit{pyopenjtalk} in the JSUT dataset and \textit{pycantonese} in the Common Voice (HK) dataset. More detailed information can be found in Appendix~\ref{info_g2p}}. We measure objective phoneme error rate (PER-O), subjective phoneme error rate (PER-S), and subjective sentence error rate (SER-S) in the evaluations. The phoneme labels in the objective PER evaluation are from the corresponding dataset (since the Common Voice (HK) dataset does not have phoneme labels, we only evaluate the subjective metrics). In the subjective evaluations, each audio in the test set is listened by at least 4 language experts. We ask them to write down the mispronounced phonemes and discuss them with each other until a conclusion is reached. Note that for SER-S, the error rate is calculated in sentence level (e.g., a sentence with multiple errors will be counted only once). More details about these evaluations can be found in Appendix~\ref{details_subjective_evaluation}. The results are shown in Table \ref{table_1}. It can be seen that Dict-TTS greatly surpasses the character-based baseline in three languages. Moreover, Dict-TTS achieves the comparable objective and subjective PER with the most popular grapheme-to-phoneme tools on two relatively larger datasets (Biaobei and Common Voice (HK)) and maintain a good pronunciation accuracy on a relatively small dataset (JSUT), which demonstrates the superiority of the explicit semantics matching in our S2PA module.

\begin{wraptable}{r}{7.8cm}\small
  \caption{The objective and subjective pronunciation accuracy comparisons in the Biaobei dataset.}
  \label{table_2}
  \centering
  \begin{tabular}{l|c|c|c}
    \toprule
    Method           & PER-O  & PER-S & SER-S        \\
    \midrule
    Character          & -      & 3.73\%  & 30.50\%     \\
    BERT Embedding~\cite{hayashi2019pre}     & -       & 4.03\% & 38.75\%    \\
    NLR~\cite{he2021neural} & -       & 2.98\% & 26.50\% \\ \midrule
    Phoneme (G2PM)     & 3.95\% & 1.39\%  & 10.50\%           \\
    Phoneme (pypinyin) & 2.78\% & 1.14\%  & 7.00\%   \\ \midrule
    Dict-TTS           & 2.12\% & 1.08\% & 6.50\%  \\
    Dict-TTS (pre-trained) & \textbf{1.54\%} &  \textbf{0.79\%} & \textbf{4.25\%} \\ 
    \bottomrule
  \end{tabular}
\end{wraptable}
We also compare the pronunciation accuracy of our Dict-TTS with various types of systems, including: 1) a character-based system; 2) a BERT embedding based system~\cite{hayashi2019pre}, where the BERT derived embeddings are concatenated with the character embeddings; 3) NLR~\cite{he2021neural}, a concurrent work that is also proposed to resolve polyphones in end-to-end TTS by extracting implicit pronunciation information from relevant dictionary texts encoded by XLM-R; 4) Phoneme (G2PM~\cite{park2020g2pm}), PortaSpeech with phoneme labels derived from G2PM (a powerful neural G2P system); 5) Phoneme (pypinyin), PortaSpeech with phoneme labels derived from pypinyin (one of the most popular Chinese G2P system). As shown in Table~\ref{table_2}, Dict-TTS greatly surpasses the systems that implicitly model the semantic representations for character-to-pronunciation mapping like NLR~\cite{he2021neural} and shows comparable performance with phoneme-based systems. Since our Dict-TTS does not require any phoneme labels for training, we can pre-train Dict-TTS on a large-scaled ASR dataset~\cite{zhang2022wenetspeech} with a small amount of effort to improve its generalization capacity. It can be seen that the pronunciation accuracy of Dict-TTS on the Biaobei dataset is significantly improved by pre-training, which demonstrates the effectiveness of our unsupervised polyphone disambiguation framework.

\begin{figure}[tbp]
	\centering
	\begin{minipage}{0.32\linewidth}
		\centering
		\includegraphics[width=1\linewidth]{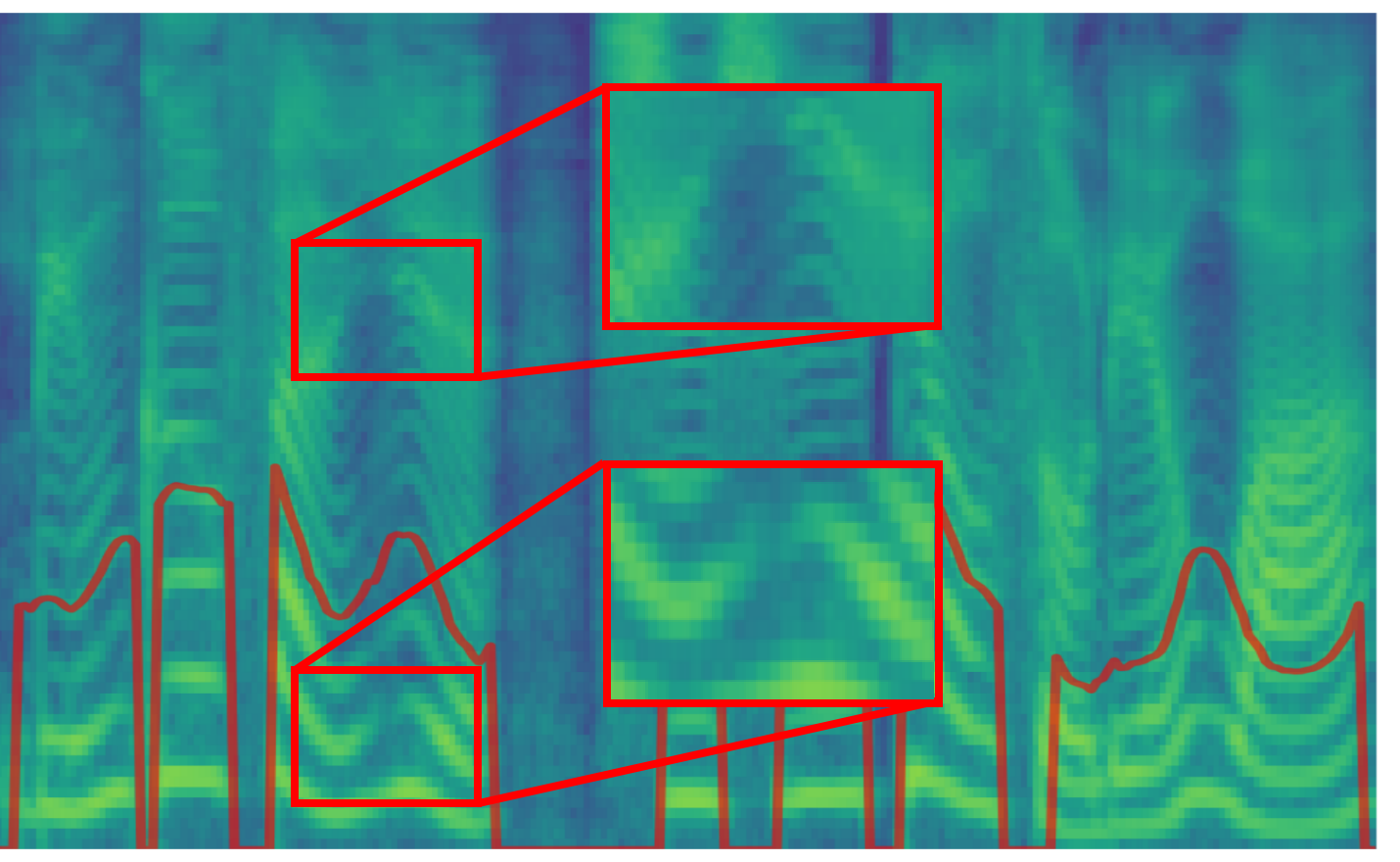}
		\caption*{(a) GT}
	\end{minipage}
	\centering
	\begin{minipage}{0.32\linewidth}
		\centering
		\includegraphics[width=1\linewidth]{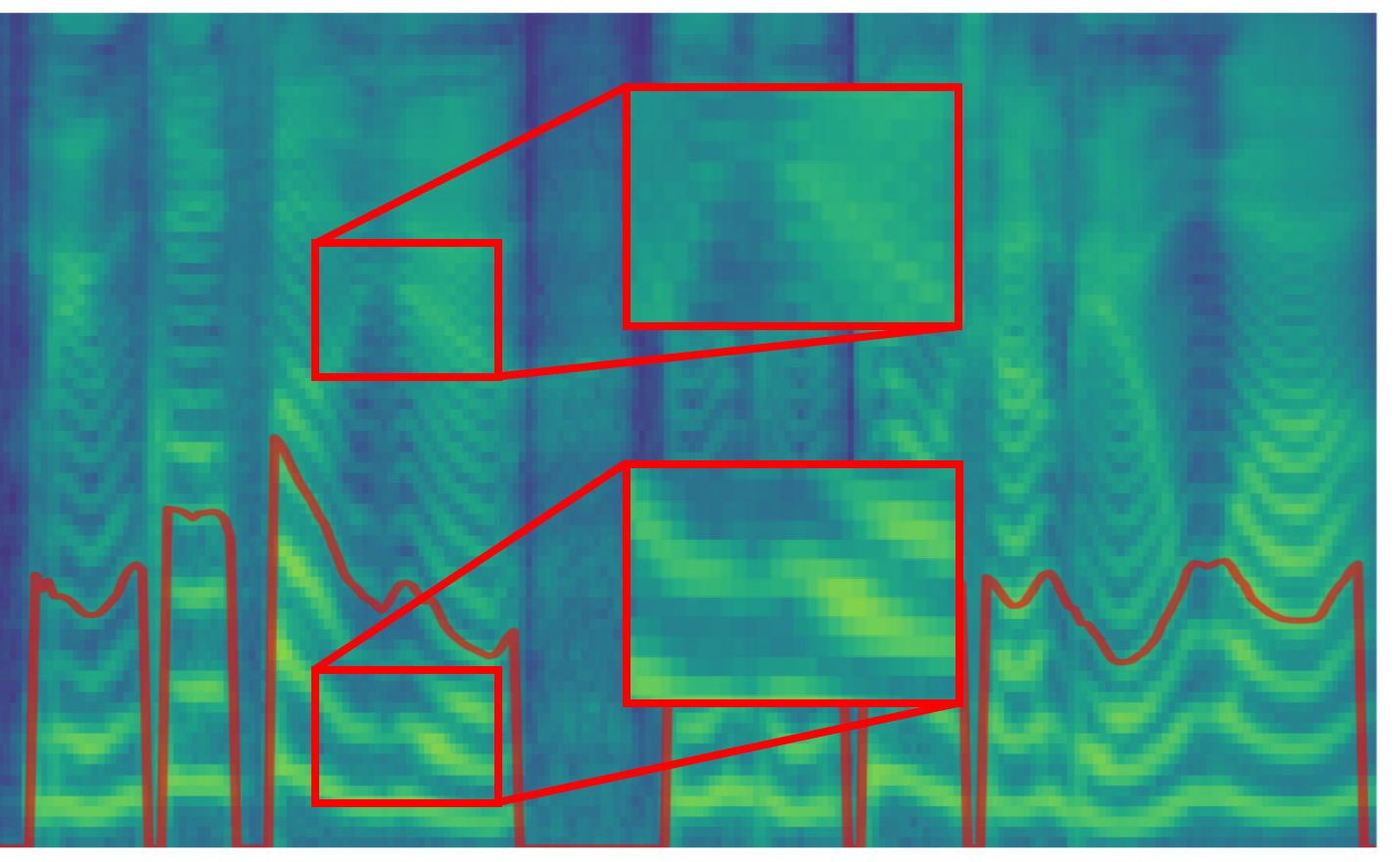}
		\caption*{(b) Character}
	\end{minipage}
	\centering
	\begin{minipage}{0.32\linewidth}
		\centering
		\includegraphics[width=1\linewidth]{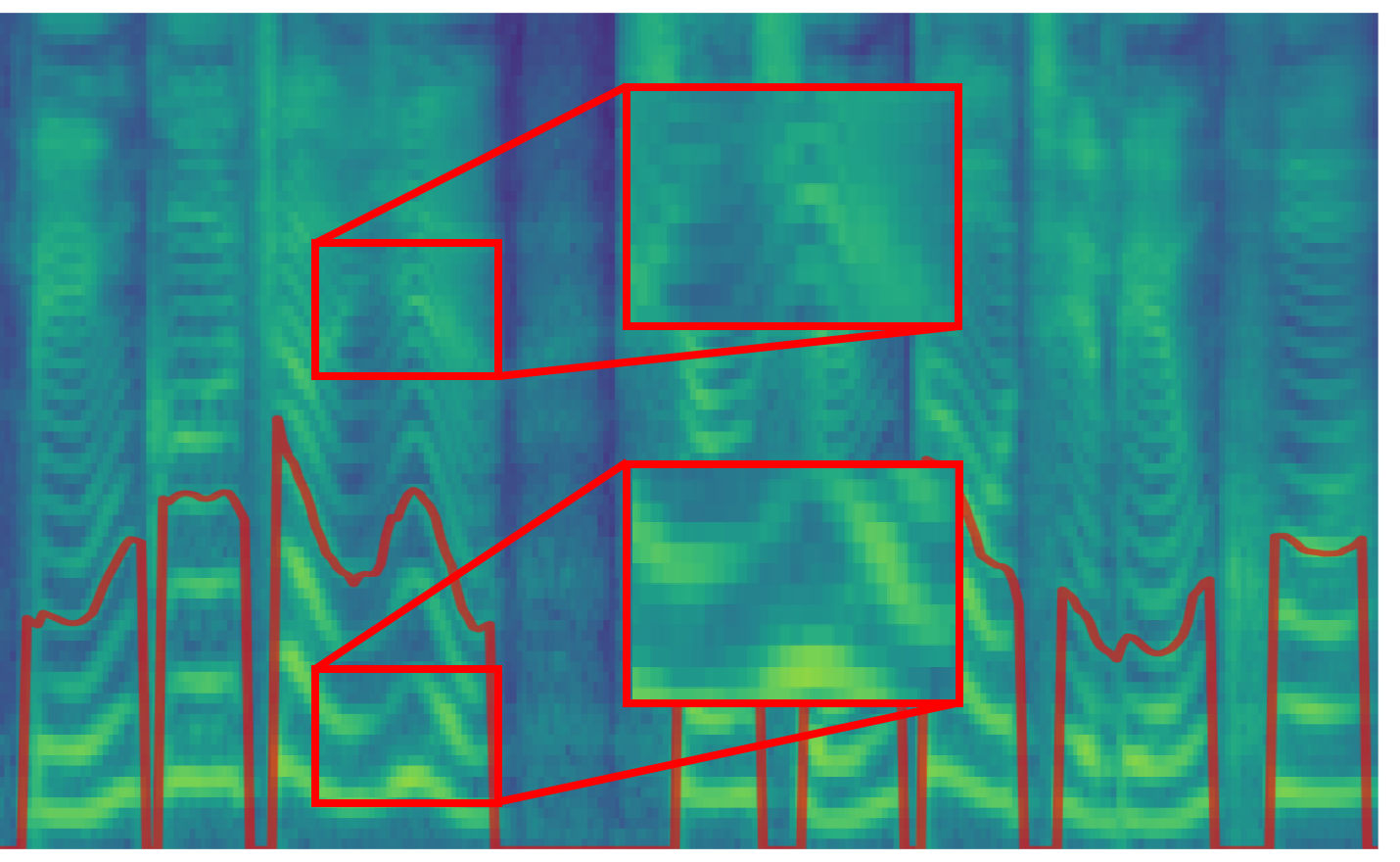}
		\caption*{(c) BERT Embedding}
	\end{minipage}
	\centering
	\begin{minipage}{0.32\linewidth}
		\centering
		\includegraphics[width=1\linewidth]{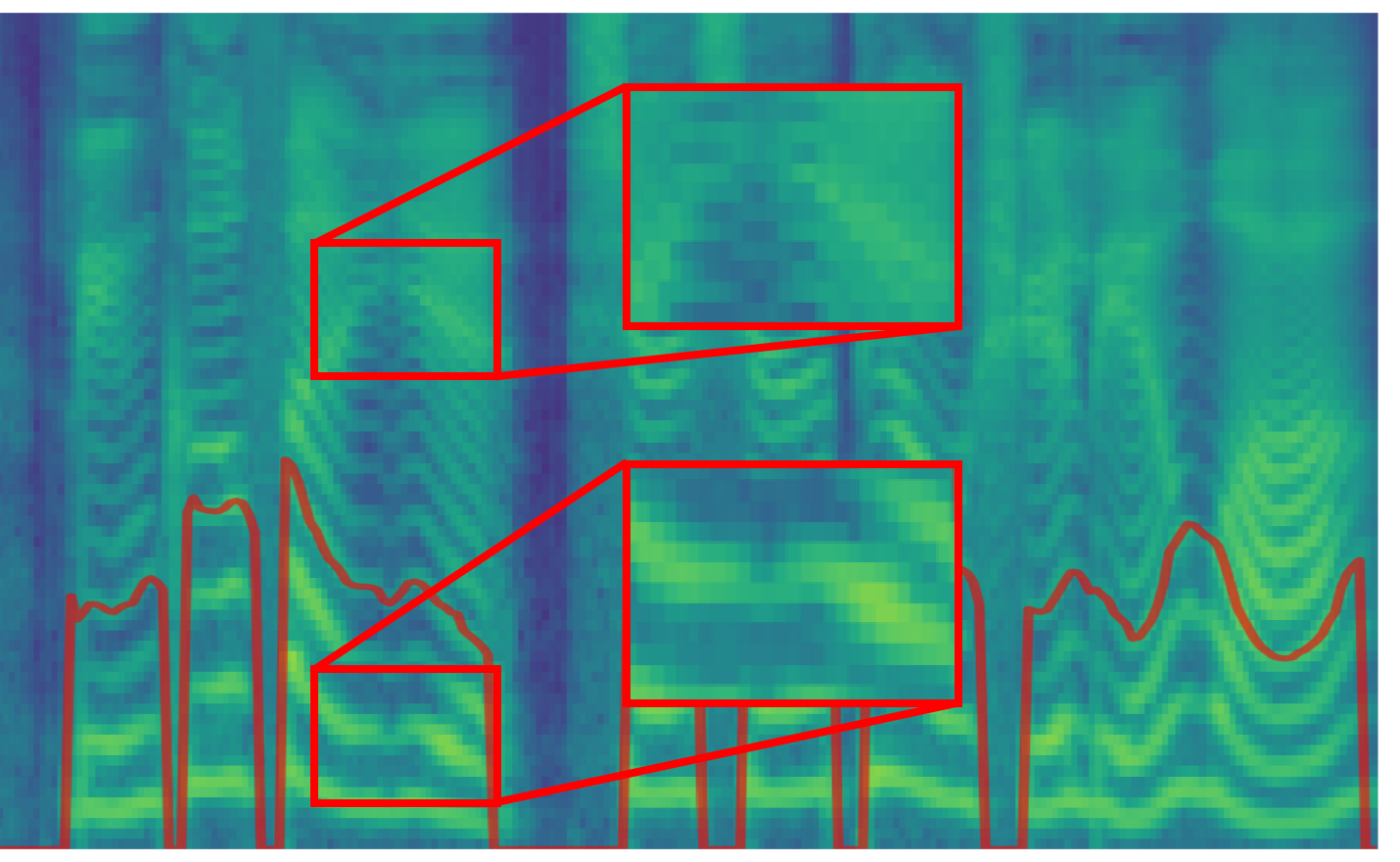}
		\caption*{(d) NLR}
	\end{minipage}
	\centering
	\begin{minipage}{0.32\linewidth}
		\centering
		\includegraphics[width=1\linewidth]{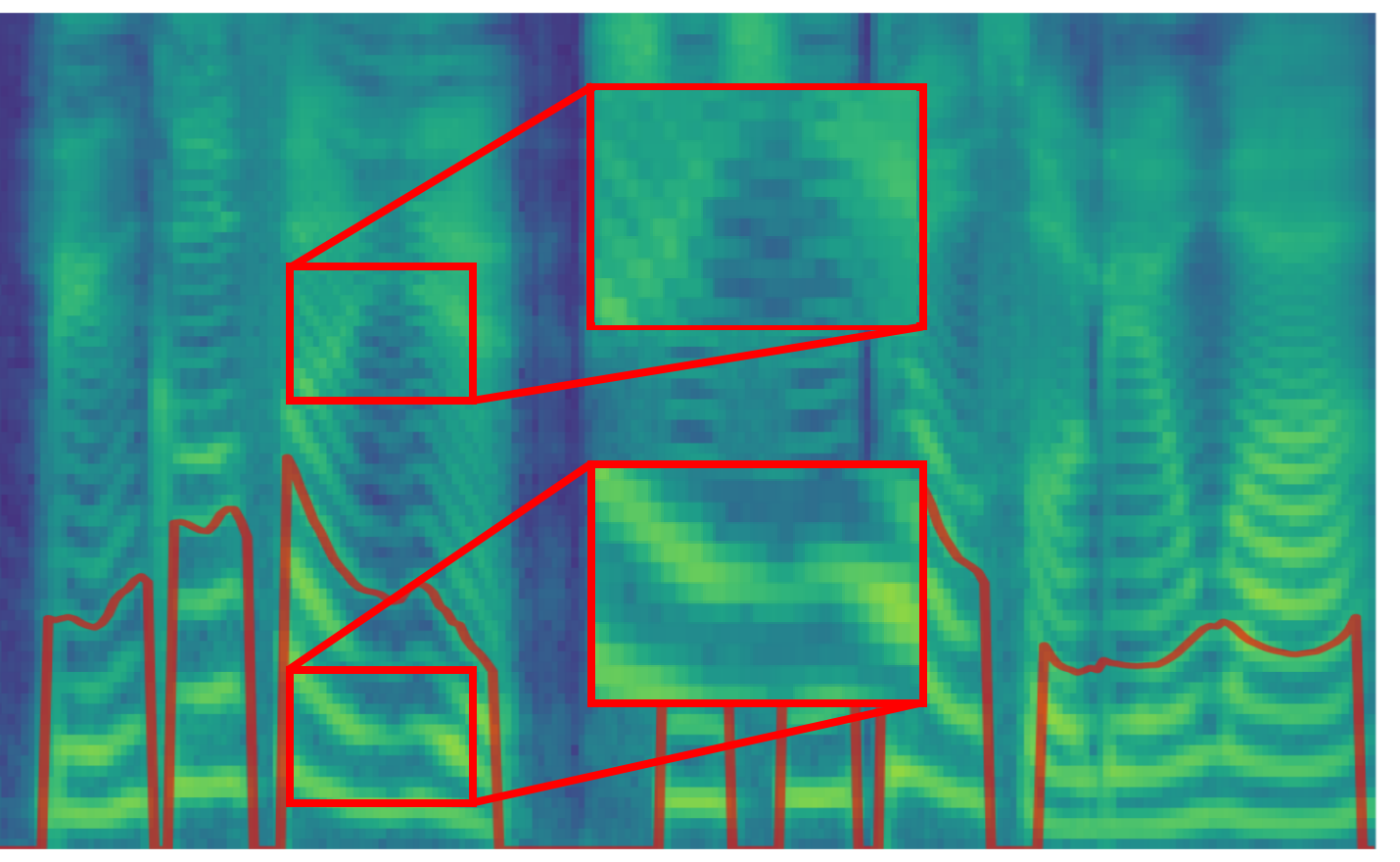}
		\caption*{(e) Phoneme (pypinyin)}
	\end{minipage}
	\centering
	\begin{minipage}{0.32\linewidth}
		\centering
		\includegraphics[width=1\linewidth]{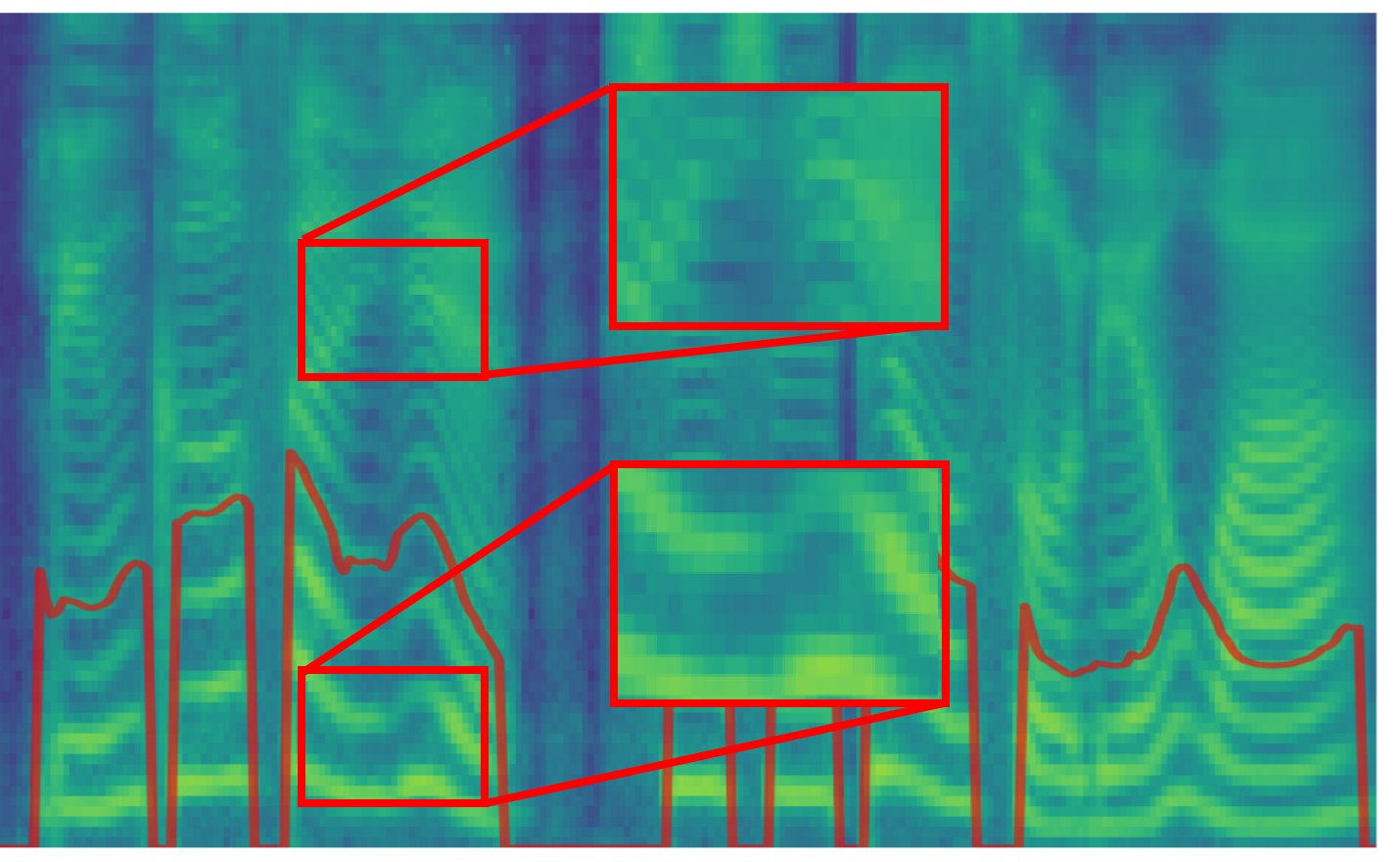}
		\caption*{(f) Dict-TTS}
	\end{minipage}
	\centering
	\caption{Visualizations of the ground-truth and generated mel-spectrograms by different types of linguistic encoders. The corresponding text is \begin{CJK}{UTF8}{gbsn}``菱歌泛夜，嬉嬉钓叟莲娃''\end{CJK}, which means ``The man picking water chestnut sings at night. The old man fishing and the girl picking lotus are laughing''.}
	\label{vis_spec_1}
\end{figure}

\subsection{Results of Audio Quality and Prosody}
We compare the audio quality, audio prosody, and pitch accuracy\footnote{We compute the average dynamic time warping (DTW)~\cite{muller2007dynamic} distances between the pitch contours of ground-truth speech and synthesized speech.} of Dict-TTS with the systems evaluated in Subsection~\ref{evaluation_pronunciation}. GT (the ground truth audio) and GT (voc.) (the ground truth audio that is firstly converted into mel-spectrogram and converted back to audio waveform using Hifi-GAN~\cite{kong2020hifi}) are also included in this experiment. We keep the text content consistent among different models to exclude other interference factors, only examining the audio quality or prosody. Each audio is listened by at least 20 testers. For audio quality and prosody, we conduct the mean opinion score (MOS) evaluation via Amazon Mechanical Turk. \begin{wraptable}{r}{7.8cm}\small
\caption{The audio performance (MOS-Q and MOS-P) and pitch accuracy comparisons. DTW denotes average dynamic time warping distances of pitch in ground-truth and synthesized audio. The mel-spectrograms are converted to waveforms using Hifi-GAN (V3)~\cite{kong2020hifi}.}
\label{table_3}
\centering
\begin{tabular}{@{}l|c|c|c@{}}
\toprule
Method    & MOS-P & MOS-Q & DTW \\ \midrule
GT        & 4.48$\pm$0.03 & 4.40$\pm$0.04 & -     \\
GT (voc.) & 4.37$\pm$0.04 & 4.26$\pm$0.04 & - \\ \midrule
Character & 3.82$\pm$0.08 & 3.88$\pm$0.07 & 53.1 \\
BERT Embedding~\cite{hayashi2019pre} & 3.88$\pm$0.07 & 3.63$\pm$0.10& 55.0 \\
NLR~\cite{he2021neural}       & 3.83$\pm$0.08 & 3.74$\pm$0.08 & 53.3 \\  \midrule
Phoneme (G2PM)    & 3.87$\pm$0.08 & 3.90$\pm$0.06& 53.1 \\
Phoneme (pypinyin)& 3.89$\pm$0.08 & \textbf{3.95$\pm$0.06} & 52.6 \\  \midrule
Dict-TTS  & \textbf{4.03$\pm$0.05} & 3.91$\pm$0.04 & \textbf{52.4} \\ \bottomrule
\end{tabular}
\end{wraptable}We analyze the MOS in two aspects: MOS-P (Prosody: naturalness of pitch, energy, and duration) and MOS-Q (Quality: clarity, high-frequency, and original timbre reconstruction). We tell the tester to focus on one corresponding aspect and ignore the other aspect when scoring. We put more information about the subjective evaluation in Appendix~\ref{details_subjective_evaluation}. As shown in Table~\ref{table_3}, for audio quality, Dict-TTS significantly outperforms those TTS systems without phoneme labels and achieves a comparable performance with phoneme-based systems. And for audio prosody and pitch accuracy, Dict-TTS even surpasses the phoneme-based systems, which demonstrates the effectiveness of the extracted semantics from prior dictionary knowledge. We put more analyses on the naturalness of prosody in Appendix~\ref{analyses_dur_pitch}. 

We then visualize the mel-spectrograms generated by the above systems in Figure~\ref{vis_spec_1}. We can see that Dict-TTS can generate mel-spectrograms with comparable details in harmonics, unvoiced frames, and high-frequency parts with the phoneme-based system, which results in similar natural sounds. Moreover, our Dict-TTS can capture more accurate local changes in pitch and speaking duration, indicating the effectiveness of introducing semantic representations in the dictionary.

\begin{table} \small
  \caption{Pronunciation accuracy, audio prosody and audio quality comparisons for ablation study.}
  \label{table_4}
  \centering
  \begin{tabular}{l|c|c|c|c|c}
    \toprule
    Settings        & PER-O  & PER-S & SER-S & CMOS-P & CMOS-Q \\
    \midrule
    Dict-TTS        & 2.12\% & 1.08\% & 6.50\% & 0.000 & 0.000 \\         
    \  w/o Semantics     & 2.15\% & 1.13\% & 6.50\% & -0.280 & -0.135 \\
    \  w/o Gumbel-Softmax     & - & 1.19\% & 7.75\% & -0.014 & -0.021 \\

    \bottomrule
  \end{tabular}
\end{table}
\subsection{Ablation Studies}
We conduct ablation studies to demonstrate the effectiveness of designs in Dict-TTS, including the auxiliary semantic information and the Gumbel-Softmax sample strategy. We conduct pronunciation accuracy and CMOS (comparative mean opinion score) evaluations for these ablation studies. The results are shown in Table~\ref{table_4}. We can see that CMOS-P drops when we remove the introduced semantic embeddings in Dict-TTS, indicating that the semantic information extracted from the dictionary can improve the audio prosody. Besides, to demonstrate the effectiveness of the Gumbel-Softmax sample strategy, we also compare the weighted sum of the pronunciation embeddings with the Gumbel-Softmax sample strategy. Since measuring PER-O requires one-hot vectors, we do not calculate the PER-O score for the weight-sum version of Dict-TTS. The results are shown in row 3 in Table~\ref{table_4}. It can be seen that PER-S and SER-S increase when we use the weighted sum method. In the experiments, the weights of different pronunciations for some characters might be close to each other, which results in relatively worse performance in the subjective results. For example, the two pronunciations ``ZH ANG3'' and ``CH ANG2'' of the character \begin{CJK}{UTF8}{gbsn}``长''\end{CJK} might be ambiguous when their weights are close to each other (e.g., $0.6$ and $0.4$). Therefore, to accurately model the subjective pronunciations, we utilize the Gumbel-Softmax function to sample the most likely pronunciation in both training and inference stages.

%Analyze the relations between semantic and acoustic space.
%3.2还需要再修改一下，尤其是phoneme部分
%3.3最好也要再加强与示意图的联系
%可以的话Section 3，加一点 interpretation ，说Dict-TTS是一种解藕方法

\section{Conclusion}
\label{conclusion}
In this paper, we proposed Dict-TTS, an unsupervised framework for polyphone disambiguation in end-to-end text-to-speech systems. Dict-TTS uses a semantics-to-pronunciation attention (S2PA) module to explicitly extract the corresponding pronunciations and semantic information from prior dictionary knowledge. The S2PA module can be trained with the end-to-end TTS model with the guidance of mel-spectrogram reconstruction loss without phoneme labels, which significantly reduces the cost of building a polyphone disambiguation system and further enables the efficient pre-training on large-scaled ASR datasets. Our experimental results in three languages show that Dict-TTS outperforms several strong G2P baseline models in terms of pronunciation accuracy and improves the prosody modeling of the baseline TTS system. Further comprehensive ablation studies demonstrate that our S2PA module successfully decomposes the character representation, pronunciation, and semantics for character-based TTS systems. However, the dictionary knowledge does not contain the tree-structured syntactic information of the input text sequence, which also affects the prosody modeling. Moreover, since we crawl the dictionary from online websites and do not make any specific changes, the performance of Dict-TTS can be further improved by a well-designed dictionary. In the future, we will try to inject syntactic information into Dict-TTS and extend it to more languages.

\section{Acknowledgments}
\label{acknowledgments}
This work was supported in part by the National Natural Science Foundation of China \(Grant No.62072397 and No.61836002\), Zhejiang Natural Science Foundation \(LR19F020006\), Yiwise, and National Key R\&D Program of China \(Grant No.2020YFC0832505\). We appreciate the support from Mindspore, which is a new deep learning computing framework.

\newpage

\bibliographystyle{plain}
\bibliography{main}

%%%%%%%%%%%%%%%%%%%%%%%%%%%%%%%%%%%%%%%%%%%%%%%%%%%%%%%%%%%%
\section*{Checklist}

\begin{enumerate}

\item For all authors...
\begin{enumerate}
  \item Do the main claims made in the abstract and introduction accurately reflect the paper's contributions and scope?
    \answerYes{}
  \item Did you describe the limitations of your work?
    \answerYes{See Section~\ref{conclusion}.}
  \item Did you discuss any potential negative societal impacts of your work?
    \answerYes{See Appendix~\ref{Negative} in the supplementary materials.}
  \item Have you read the ethics review guidelines and ensured that your paper conforms to them?
    \answerYes{}
\end{enumerate}

\item If you are including theoretical results...
\begin{enumerate}
  \item Did you state the full set of assumptions of all theoretical results?
    \answerNA{}
	\item Did you include complete proofs of all theoretical results?
    \answerNA{}
\end{enumerate}

\item If you ran experiments...
\begin{enumerate}
  \item Did you include the code, data, and instructions needed to reproduce the main experimental results (either in the supplemental material or as a URL)?
    \answerYes{See supplementary materials.}
  \item Did you specify all the training details (e.g., data splits, hyperparameters, how they were chosen)?
    \answerYes{See Subsection~\ref{Experimental Setup}.}
	\item Did you report error bars (e.g., with respect to the random seed after running experiments multiple times)?
    \answerYes{We report confidence intervals of subjective metric results and describe the random seed settings in Appendix in the supplementary materials.}
	\item Did you include the total amount of compute and the type of resources used (e.g., type of GPUs, internal cluster, or cloud provider)?
    \answerYes{See Subsection~\ref{Implementation Details}.}
\end{enumerate}

\item If you are using existing assets (e.g., code, data, models) or curating/releasing new assets...
\begin{enumerate}
  \item If your work uses existing assets, did you cite the creators?
    \answerYes{See Subsection~\ref{Implementation Details}.}
  \item Did you mention the license of the assets?
    \answerNA{}
  \item Did you include any new assets either in the supplemental material or as a URL?
    \answerNA{}
  \item Did you discuss whether and how consent was obtained from people whose data you're using/curating?
    \answerNA{}
  \item Did you discuss whether the data you are using/curating contains personally identifiable information or offensive content?
    \answerNA{}
\end{enumerate}

\item If you used crowdsourcing or conducted research with human subjects...
\begin{enumerate}
  \item Did you include the full text of instructions given to participants and screenshots, if applicable?
    \answerYes{See Appendix~\ref{details_subjective_evaluation} in the supplementary materials.}
  \item Did you describe any potential participant risks, with links to Institutional Review Board (IRB) approvals, if applicable?
    \answerNA{}
  \item Did you include the estimated hourly wage paid to participants and the total amount spent on participant compensation?
    \answerYes{See Appendix~\ref{details_subjective_evaluation} in the supplementary materials.}
\end{enumerate}

\end{enumerate}

%%%%%%%%%%%%%%%%%%%%%%%%%%%%%%%%%%%%%%%%%%%%%%%%%%%%%%%%%%%%
\newpage
\appendix
\appendixpage

\section{Details of Models}
In this section, we describe details about the dictionary construction, the architecture of the semantic encoder and linguistic encoder, and multi-length discriminator. For further information about the variational generator, one could refer to PortaSpeech~\cite{ren2021portaspeech}. We also describe our modifications to model architecture to support the pre-training task of Dict-TTS with large-scaled ASR datasets.

\subsection{Dictionary Construction}
\label{dictionary_detail}
The Chinese dictionary used in the experiments is obtained from \url{https://github.com/yihui/zdict}, which is crawled from \url{https://www.zdic.net/zd/zb/ty/}. The Japanese dictionary is crawled from \url{https://dictionary.goo.ne.jp} and the Cantonese dictionary is crawled from \url{https://humanum.arts.cuhk.edu.hk/Lexis/lexi-can}. The crawled dictionaries are processed into the format described in SubSection~\ref{method_S2PA}. 

We use a pre-trained cross-lingual language model~\cite{conneau2020unsupervised} to extract the semantic context information $k$ of every gloss entry $e$ and store them in the disk for computational efficiency\footnote{The model used in our experiments can be downloaded from \url{https://huggingface.co/xlm-roberta-base}. One could use monolingual BERT to enhance the performance.}. Note that our Dict-TTS does not need the heavy BERT model in the inference stage. As is pointed out by the previous works, lower layers of BERT are found to perform broad attention across all pairs of words or encode local syntax~\cite{clark2019does,peters2018dissecting} and middle layers are found to mostly capture transferable syntactic and semantic knowledge~\cite{hewitt2019structural,tenney2019bert}. But the upper layers are specifically tuned towards the pre-training tasks of BERT~\cite{xia2021using}. Thus, we use the average of the first-layer word embeddings and the later 8-layer contextual BERT representations as the extracted knowledge.

\subsection{Architecture of the Semantic Encoder and Linguistic Encoder}
\begin{wrapfigure}[17]{r}{14em}
  \centering
    \includegraphics[scale=0.45]{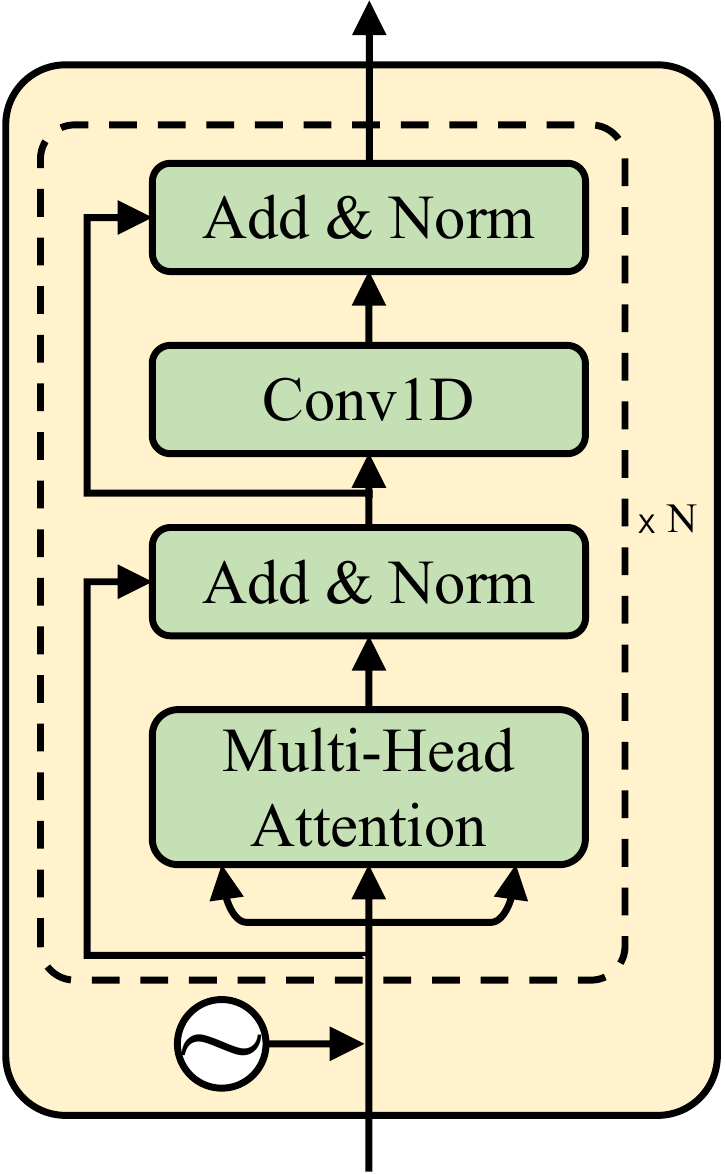}
  \caption{The detailed architecture of the semantic encoder and linguistic encoder.}
  \label{semancitc_linguistic_encoder}
\end{wrapfigure}
As shown in Figure~\ref{semancitc_linguistic_encoder}, the semantic encoder and linguistic encoder of our Dict-TTS
are both stacks of feed-forward Transformer layers with relative position encoding~\cite{shaw2018self} following PortaSpeech~\cite{ren2021portaspeech}. And there are four layers of the semantic encoder and four layers of the linguistic encoder in our Dict-TTS.

\subsection{Modifications for Pre-training Task}
The vanilla architecture of PortaSpeech is tested on LJSpeech (a single-speaker dataset). By introducing a group of learnable speaker embeddings to represent the speakers’ timbre, pronunciation habits, and other features, it can be trained on multi-speaker datasets. However, since large-scaled ASR datasets may not have explicit speaker information, the following modifications should be made: we extract the speaker embeddings from audio samples using resemblyzer\footnote{\url{https://github.com/resemble-ai/Resemblyzer}} and feed them into the variational generator and duration predictor. We pre-train our DictTTS on the WenetSpeech dataset~\cite{zhang2022wenetspeech}, which is a multi-domain corpus for speech recognition. It takes 600k steps for pre-training until convergence. As is shown in Table~\ref{table_2}, the semantic comprehension and the generalization capacity of our Dict-TTS
are significantly improved by our pre-training process.

%[8] What Does BERT Look At
%[16] A structural probe for finding syntax in word representations
%[36] BERT Rediscovers the Classical NLP Pipeline
%[at] Using Prior Knowledge to Guide BERT’s Attention in Semantic Textual Matching Tasks

\section{Detailed Experimental Settings}
In this section, we describe more model configurations and details in subjective evaluation.

\subsection{Dict-TTS Model Configurations}
\label{dicttts_config}
We list the model hyper-parameters of Dict-TTS in Table~\ref{table_5}.
\begin{table}[h]\small
\caption{Hyperparameters of Dict-TTS models.}
\label{table_5}
\centering
\begin{tabular}{@{}llcc@{}}
\toprule
\multicolumn{2}{c}{Hyper-parameter}                                                                                    & \multicolumn{1}{l}{Dict-TTS}     & \multicolumn{1}{l}{Number of parameters} \\ \midrule
\multicolumn{1}{l|}{\multirow{5}{*}{Semantic Encoder}}           & \multicolumn{1}{l|}{Character Embedding}            & \multicolumn{1}{c|}{192}         & \multirow{5}{*}{4.907M}                  \\
\multicolumn{1}{l|}{}                                            & \multicolumn{1}{l|}{Semantic Encoder Layers}        & \multicolumn{1}{c|}{4}           &                                          \\
\multicolumn{1}{l|}{}                                            & \multicolumn{1}{l|}{Hidden Size}                    & \multicolumn{1}{c|}{192}         &                                          \\
\multicolumn{1}{l|}{}                                            & \multicolumn{1}{l|}{Conv1D Kernel}                  & \multicolumn{1}{c|}{5}           &                                          \\
\multicolumn{1}{l|}{}                                            & \multicolumn{1}{l|}{Conv1D Filter Size}             & \multicolumn{1}{c|}{768}         &                                          \\ \midrule
\multicolumn{1}{l|}{S2PA Module}                                 & \multicolumn{1}{l|}{Hidden Size}                    & \multicolumn{1}{c|}{192}         & 0.404M                                   \\ \midrule
\multicolumn{1}{l|}{\multirow{4}{*}{Linguistic Encoder}}         & \multicolumn{1}{l|}{Linguistic Encoder Layers}      & \multicolumn{1}{c|}{4}           & \multirow{4}{*}{3.371M}                  \\
\multicolumn{1}{l|}{}                                            & \multicolumn{1}{l|}{Hidden Size}                    & \multicolumn{1}{c|}{192}         &                                          \\
\multicolumn{1}{l|}{}                                            & \multicolumn{1}{l|}{Conv1D Kernel}                  & \multicolumn{1}{c|}{5}           &                                          \\
\multicolumn{1}{l|}{}                                            & \multicolumn{1}{l|}{Conv1D Filter Size}             & \multicolumn{1}{c|}{768}         &                                          \\ \midrule
\multicolumn{1}{l|}{\multirow{8}{*}{Variational Generator}}      & \multicolumn{1}{l|}{Encoder Layers}                 & \multicolumn{1}{c|}{8}           & \multirow{8}{*}{7.516M}                  \\
\multicolumn{1}{l|}{}                                            & \multicolumn{1}{l|}{Decoder Layers}                 & \multicolumn{1}{c|}{4}           &                                          \\
\multicolumn{1}{l|}{}                                            & \multicolumn{1}{l|}{Encoder/Decoder Kernel}         & \multicolumn{1}{c|}{5}           &                                          \\
\multicolumn{1}{l|}{}                                            & \multicolumn{1}{l|}{Encoder/Decoder channel size}   & \multicolumn{1}{c|}{192}         &                                          \\
\multicolumn{1}{l|}{}                                            & \multicolumn{1}{l|}{Latent Size}                    & \multicolumn{1}{c|}{16}          &                                          \\
\multicolumn{1}{l|}{}                                            & \multicolumn{1}{l|}{Prior Flow Layers}              & \multicolumn{1}{c|}{4}           &                                          \\
\multicolumn{1}{l|}{}                                            & \multicolumn{1}{l|}{Prior Flow Conv1D Kernel}       & \multicolumn{1}{c|}{3}           &                                          \\
\multicolumn{1}{l|}{}                                            & \multicolumn{1}{l|}{Prior Flow Conv1D Channel Size} & \multicolumn{1}{c|}{64}          &                                          \\ \midrule
\multicolumn{1}{l|}{\multirow{4}{*}{Multi-Length Discriminator}} & \multicolumn{1}{l|}{Number of Discriminators}       & \multicolumn{1}{c|}{3}           & \multirow{4}{*}{0.927M}                  \\
\multicolumn{1}{l|}{}                                            & \multicolumn{1}{l|}{Window Size}                    & \multicolumn{1}{c|}{32, 64, 128} &                                          \\
\multicolumn{1}{l|}{}                                            & \multicolumn{1}{l|}{Conv2D Layers}                  & \multicolumn{1}{c|}{3}           &                                          \\
\multicolumn{1}{l|}{}                                            & \multicolumn{1}{l|}{Hidden Size}                    & \multicolumn{1}{c|}{192}         &                                          \\ \midrule
\multicolumn{3}{c|}{Total Number of Parameters}                                                                                                           & 17.125M                                  \\ \bottomrule
\end{tabular}
\end{table}

\subsection{Baseline Model Configurations}
\label{baseline_config}
The baseline systems in our experiments can be divided into: 1) phoneme-based systems; 2) character-based systems. For phoneme-based systems, we use the mixture alignment proposed in PortaSpeech~\cite{ren2021portaspeech}, which takes phoneme sequence as inputs but utilize both soft phoneme-level and hard word-level duration for mixture alignment. We use a 4-layer phoneme encoder and a 4-layer word encoder following the implementation of PortaSpeech~\cite{ren2021portaspeech}. For character-based systems, we use an 8-layer character encoder for fair comparisons. The encoders above are stacks of feed-forward Transformer layers with relative position encoding~\cite{shaw2018self} following PortaSpeech~\cite{ren2021portaspeech}. Other parts of the architecture are kept the same as Dict-TTS to exclude other interference factors.

\begin{figure}[tbp]
    \centering
	\begin{minipage}{0.85\linewidth}
		\centering
		\includegraphics[width=1\linewidth]{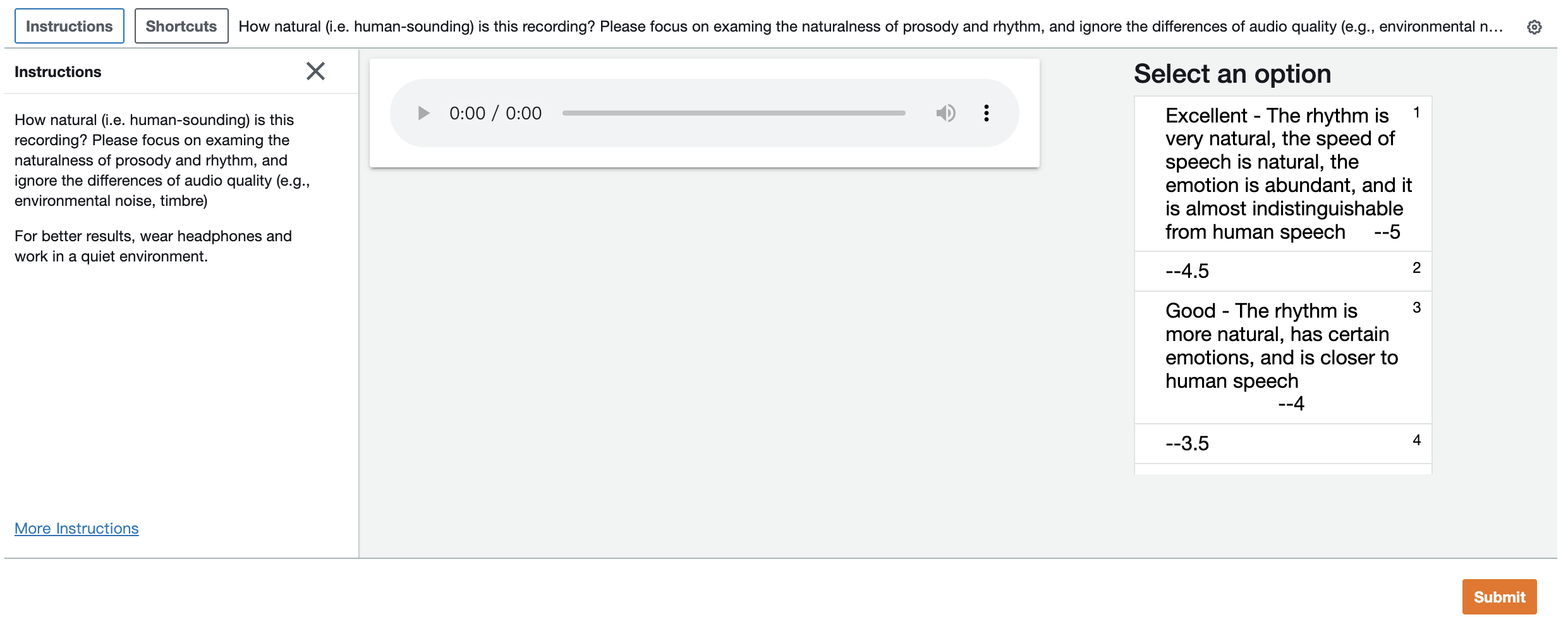}
		\caption*{(a)  Screenshot of MOS-P testing.}
	\end{minipage}
	\centering
	\begin{minipage}{0.85\linewidth}
		\centering
		\includegraphics[width=1\linewidth]{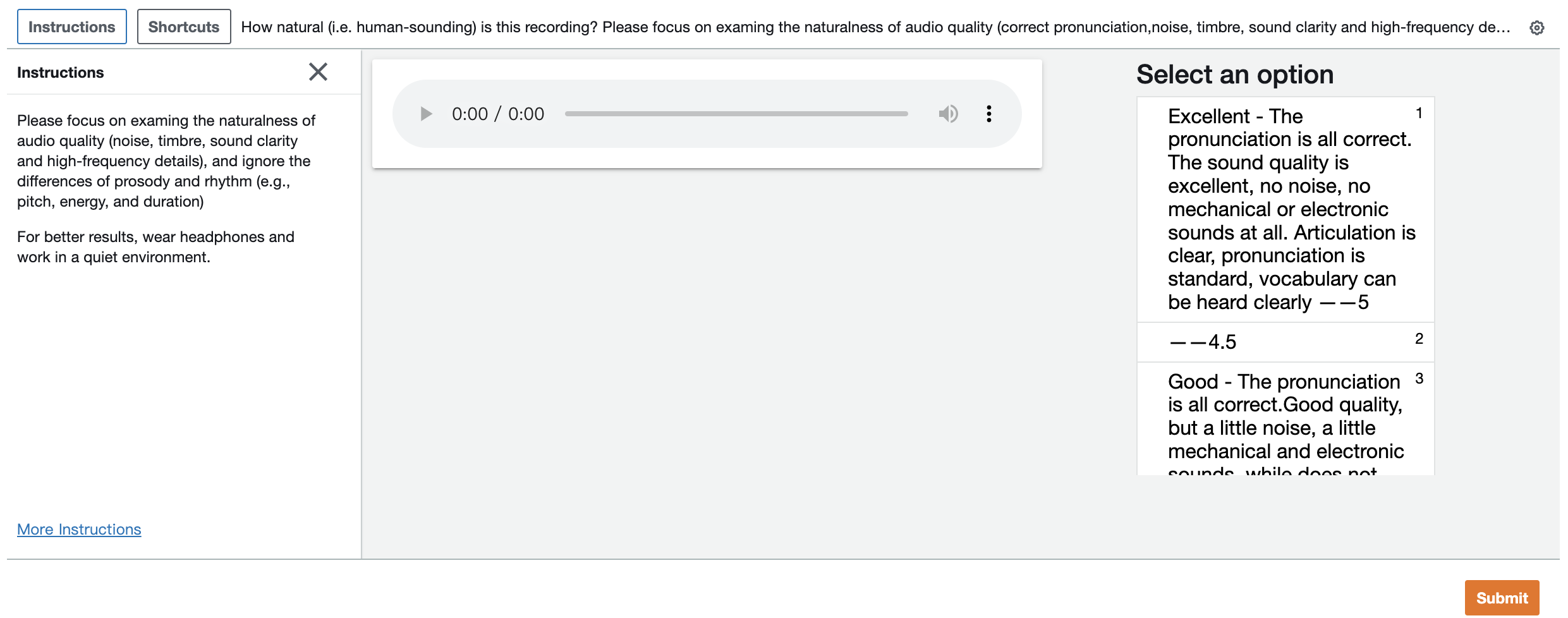}
		\caption*{(b) Screenshot of MOS-Q testing.}
	\end{minipage}
	\centering
	\begin{minipage}{0.85\linewidth}
		\centering
		\includegraphics[width=1\linewidth]{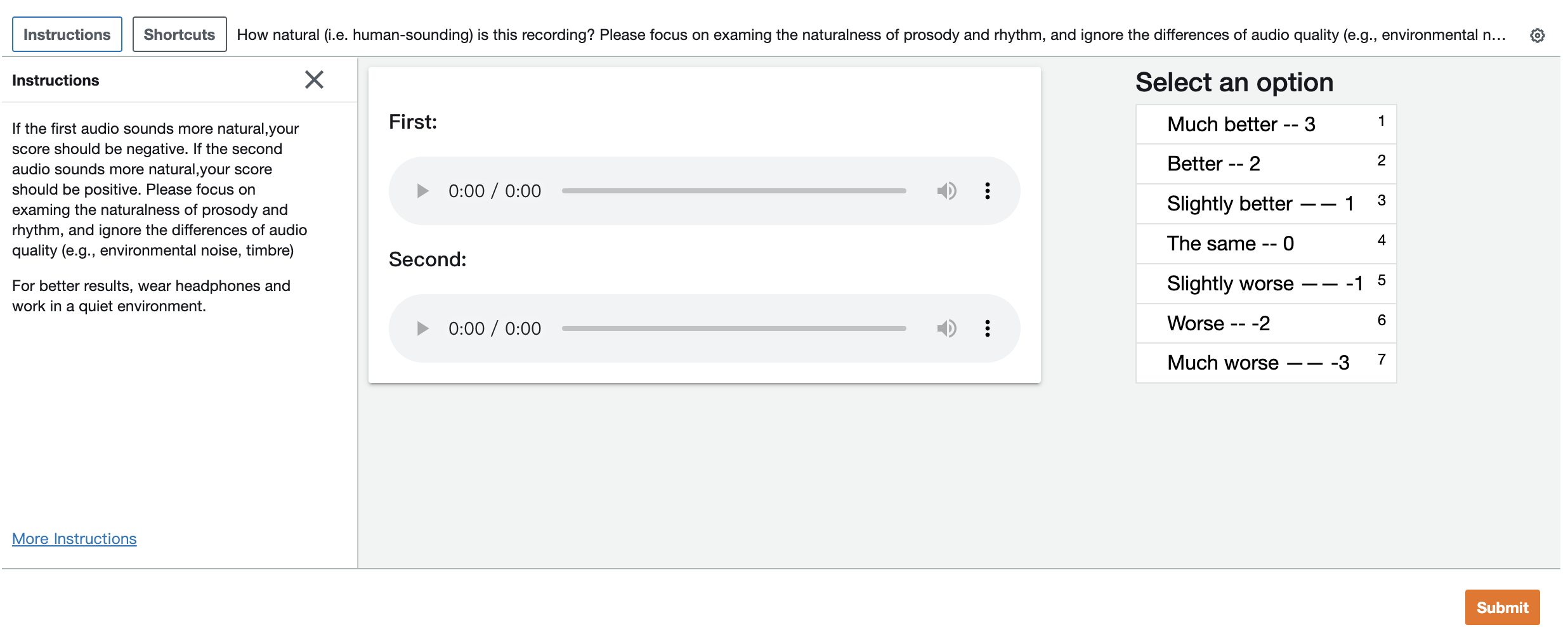}
		\caption*{(c) Screenshot of CMOS-P testing.}
	\end{minipage}
	\centering
	\begin{minipage}{0.85\linewidth}
		\centering
		\includegraphics[width=1\linewidth]{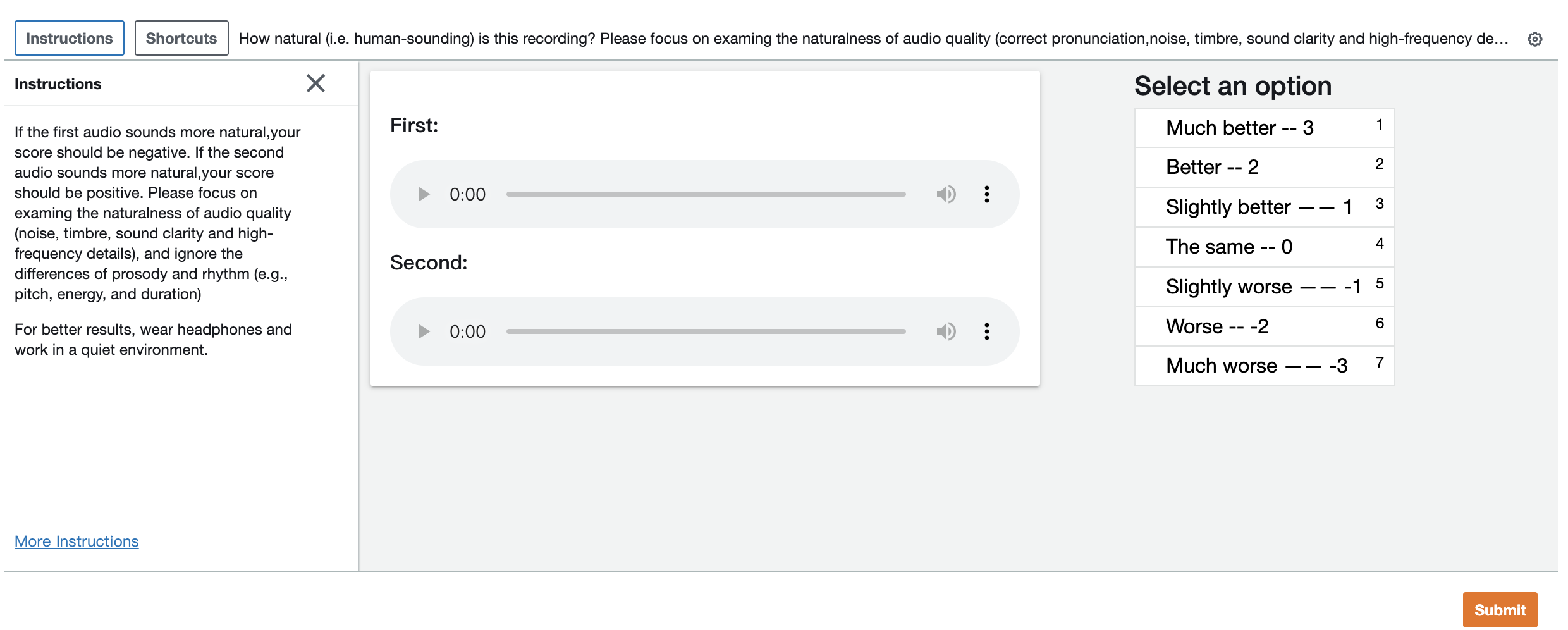}
		\caption*{(d) Screenshot of CMOS-Q testing.}
	\end{minipage}
	\centering
	\caption{Screenshots of audio quality and prosody evaluations.}
	\label{screenshots_audio_quality_prosody}
\end{figure}

\begin{figure}[tbp]
    \centering
    \includegraphics[width=1\linewidth]{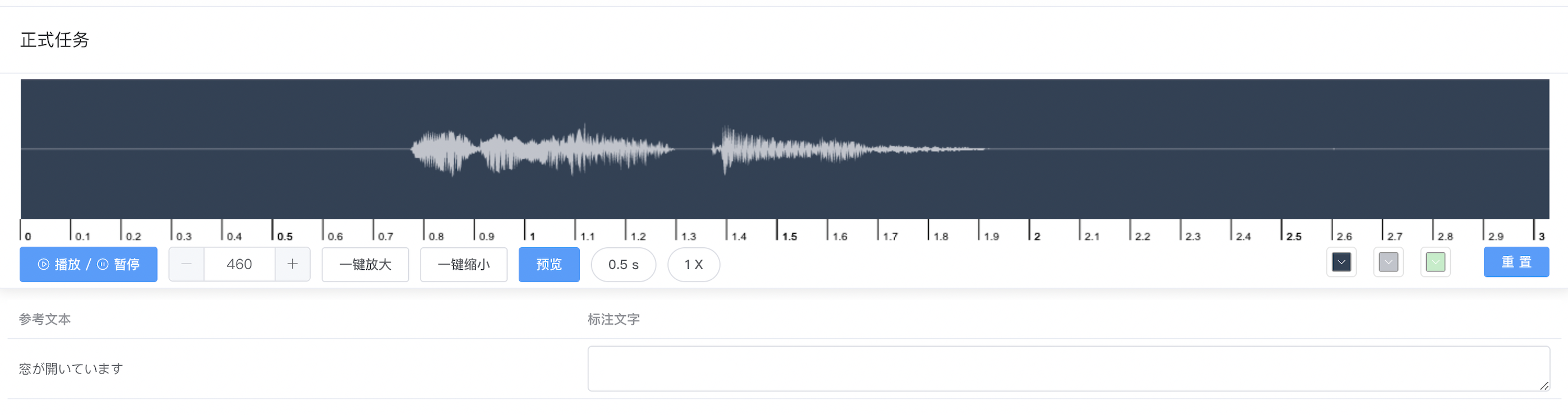}
	\caption{Screenshot of pronunciation accuracy evaluation.}
	\label{screenshots_pronunciation_accuracy}
\end{figure}

\subsection{Details in Subjective Evaluation}
\label{details_subjective_evaluation}
\paragraph{Audio Quality and Prosody} We perform the audio quality and prosody evaluation on Amazon Mechanical Turk (MTurk). For each dataset, we randomly select 50 texts from the test set and use the TTS systems to generate the audio samples. Each audio has been listened to by at least 20 listeners. For MOS, each tester is asked to evaluate the subjective naturalness of a sentence on a 1-5 Likert scale. For CMOS, listeners are asked to compare pairs of audio generated by systems A and B, indicate which of the two audio they prefer, and choose one of the following scores: 0 indicating no difference, 1 indicating a slight difference, 2 indicating a significant difference and 3 indicating a very large difference. For audio quality evaluation (MOS-Q and CMOS-Q), we tell listeners to ``\textit{focus on examining the naturalness of audio quality (e.g., noise, timbre, sound clarity, and high-frequency details), and ignore the differences of prosody and rhythm (e.g., pitch, energy, and duration)}''. For prosody evaluations (MOS-P and CMOS-P), we tell listeners to ``\textit{focus on examining the naturalness of prosody and rhythm (e.g., pitch, energy, and duration), and ignore the differences in audio quality (e.g., noise, timbre, sound clarity, and high-frequency details)}''. The screenshots of instructions for testers are shown in Figure~\ref{screenshots_audio_quality_prosody}. We paid \$8 to participants hourly and totally spent about \$500 on participant compensation.

\paragraph{Pronunciation Accuracy} We perform the pronunciation accuracy evaluation on MolarData\footnote{\url{https://www.molardata.com/en.html}}. For each dataset, we use all texts in the test set to generate the audio samples. Each audio has been listened to by at least 4 language experts. Each tester is asked to carefully listen to the audio multiple times, write down the mispronounced phonemes, and discuss with each other until a conclusion is reached. The screenshots of instructions for testers are shown in Figure~\ref{screenshots_pronunciation_accuracy}. We paid \$15 to participants hourly and totally spent about \$800 on participant compensation. A small subset of speech samples used in the test is available at \url{https://dicttts.github.io/DictTTS-Demo/}.

\subsection{Details of the G2P tools used in the experiments}
\label{info_g2p}
We use \textit{pypinyin (0.46.0)}\footnote{\url{https://github.com/mozillazg/python-pinyin}} and G2PM\footnote{\url{https://github.com/kakaobrain/g2pM}} in Biaobei dataset. Until 20 May, 2022, the latest version of pypinyin is 0.46.0. And it is worth noting that the pronunciation accuracy is greatly improved when we update the version from 0.36.0 to 0.46.0. In JSUT dataset, we use \textit{pyopenjtalk (0.2.0)}\footnote{\url{https://github.com/r9y9/pyopenjtalk}}. And in Common Voice (HK) dataset, we use \textit{pycantonese (3.4.0)}\footnote{\url{https://github.com/jacksonllee/pycantonese}}.

\subsection{Details of Datasets}
\begin{comment}
\begin{table}[hp]
\caption{The number of polyphones in every dataset by the number of possible pronunciations}
\label{table datasets}
\centering
\begin{tabular}{@{}cllll@{}}
\toprule
Pronunciations  & \multicolumn{1}{c}{Biaobei} & \multicolumn{1}{c}{Common voice HK} & \multicolumn{1}{c}{JUST} & \multicolumn{1}{c}{AISHEEL-3} \\ \midrule
1               & 1221(46.2\%)                & 2348(63.4\%)                        & 920(36.5\%)              & 945(37.7\%)                   \\
2               & 887(33.6\%)                 & 957(25.9\%)                         & 793(31.5\%)              & 975(38.8\%)                   \\
3               & 356(13.4\%)                 & 280(7.6\%)                          & 442(17.5\%)              & 397(15.8\%)                   \\
\textgreater{}3 & 178(6.8\%)                  & 115(3.1\%)                          & 364(14.5\%)              & 193(7.7\%)                    \\
Total           & 2642(100\%)                 & 3700(100\%)                         & 2519(100\%)              & 2510(100\%)                   \\
\textgreater{}1 & 1421(53.8\%)                & 1352(36.6\%)                        & 1599(63.5\%)             & 1565(62.3\%)                  \\ \bottomrule
\end{tabular}
\end{table}
\end{comment}

\begin{figure}[hp]
  \centering
    \includegraphics[scale=0.48]{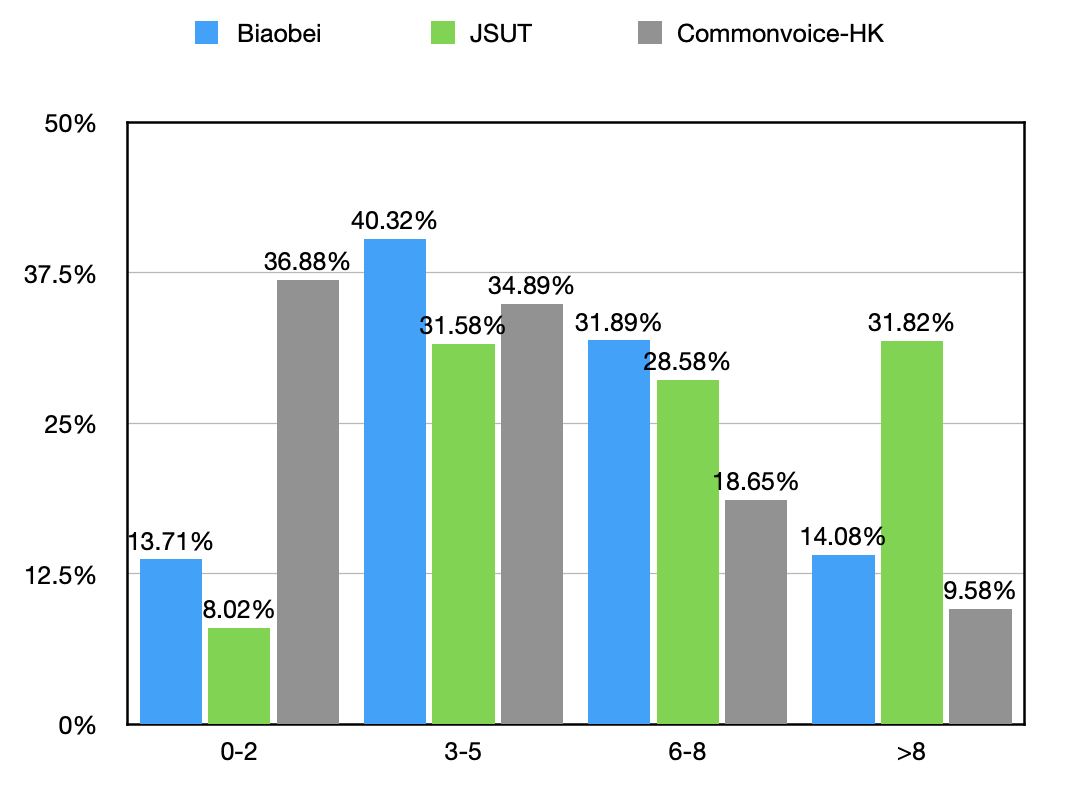}
  \caption{The illustration of the number of polyphone in a sentence. The horizontal axis is the frequency of polyphones in a sentence and the vertical axis is the number of sentences shown by percentage.}
  \label{figure datasets}
\end{figure}
In this subsection, we mainly describe the distribution characteristics of polyphones in the datasets used in our experiments, including: 1) Biaobei~\cite{baker2017chinese}; 2) JSUT~\cite{sonobe2017jsut} and 3) Common Voice (HK)~\cite{ardila2020common}. As shown in Figure~\ref{figure datasets}, we calculate the number of polyphones in every sentence. The average sentence lengths of the Biaobei, JSUT, and Common Voice (HK) datasets are 18, 29, and 10. Among these datasets, the most frequent number of polyphones appearing in a sentence is three to five and some sentences even include eight (or more) polyphones, indicating the importance of polyphone disambiguation.

%Table~\ref{table datasets} shows the number of polyphones in every dataset, divided by the number of possible pronunciations. Note that we ignore the repetition of the same word in different sentence. Among all the datasets, Cantonese speech corpus Commonvoice-HK owns the least number of polyphone in all words, which is 1352(36.6\%) , while Japanese speech corpus JUST owns the largest number of polyphone, which is 1599(63.5\%). The number of polyphones in Chinese speech corpus Biaobei and AISHELL-3 are 1421(53.8\%) and 1565(62.3\%), separately. We also present the explicit number of polyphones in each dataset , proving that 1)Japanese polyphones do has more pronunciations among the other(14.5\% of words has more than three pronunciations in JUST); 2)Chinese and Cantonese dataset Biaobei, AISHELL-3 and Commonvoice-HK are more concentration of two pronunciations polyphones(33.6\% in Biaobei, 38.8\% in AISHELL-3, 25.9\% in Commonvoice-HK). 

\subsection{Error Bars and Random Seeds}
For the experiments of the audio quality and prosody, we report confidence intervals of subjective metric results in Table~\ref{table_3}. For the experiments of the pronunciation accuracy, we ran the experiments 10 times with 10 different random seeds ($[1234,1111,2222,3333,4444,5555,6666,7777,8888,9999]$) and obtained the averaged results.

\begin{figure}[tbp]
    \centering
	\begin{minipage}{0.95\linewidth}
		\centering
		\includegraphics[width=1\linewidth]{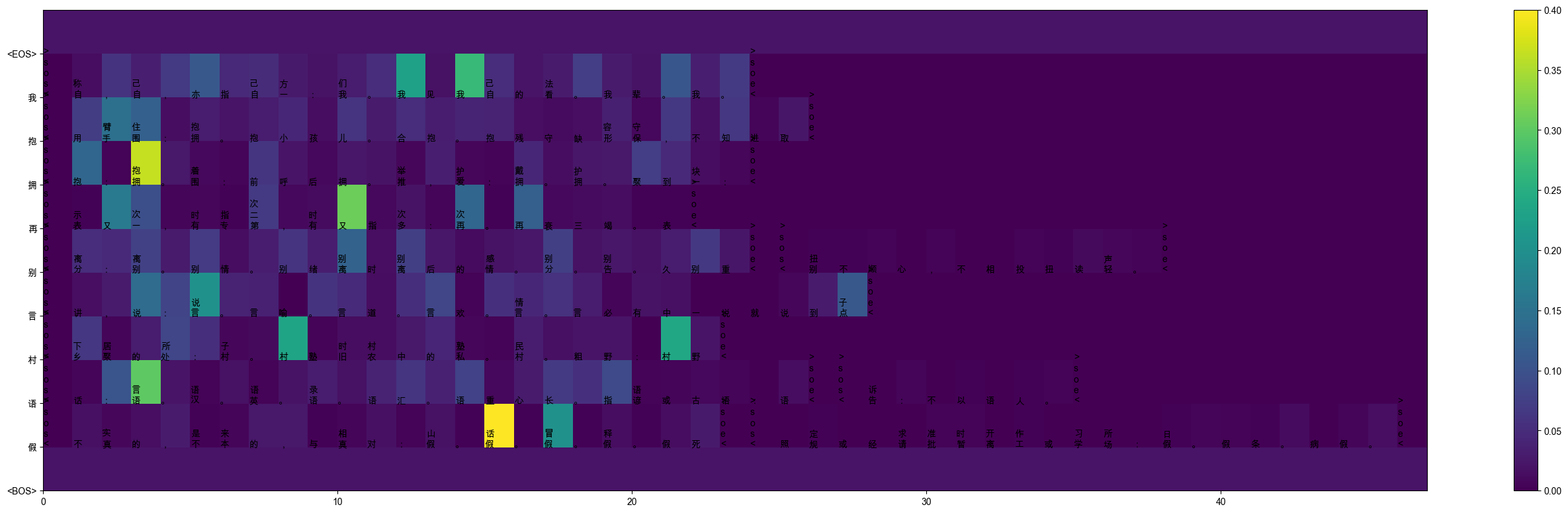}
		\caption*{(a) Biaobei-000001}
	\end{minipage}
	\centering
	\begin{minipage}{0.95\linewidth}
		\centering
		\includegraphics[width=1\linewidth]{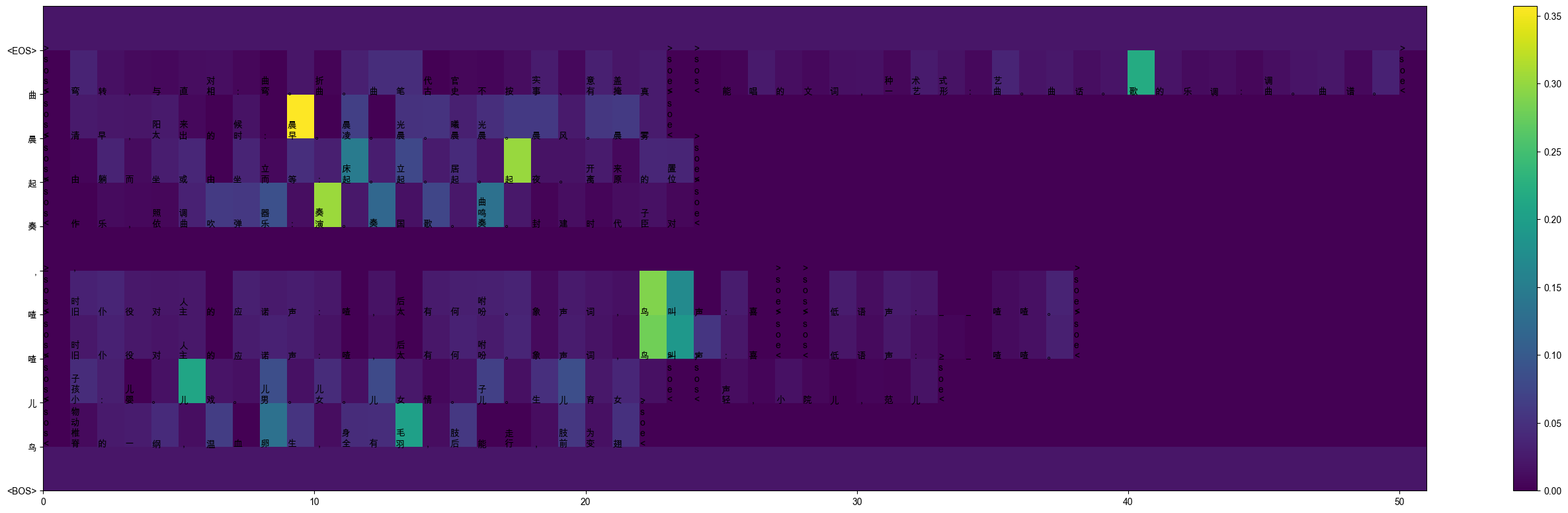}
		\caption*{(b) Biaobei-000002}
	\end{minipage}
	\centering
	\caption{Visualizations of the semantics attention weights in Biaobei dataset.}
	\label{visualization_attn_weights}
\end{figure}
\section{Visualization of Attention Weights}
We put some semantics attention visualizations in Figure~\ref{visualization_attn_weights}. We can see that Dict-TTS can create reasonable text-to-dictionary alignments, which improves the performance of polyphone disambiguation and helps the semantics comprehension for end-to-end TTS systems.

\section{The Definition of Polyphone and Heteronym}
\label{definition_hp}
Polyphones are letters or characters having more than one phonetic value, while heteronyms are words that have two (or more) different possible pronunciations that are associated with two (or more) different meanings~\cite{martin1981heteronyms}. According to the previous research~\cite{zhang2020mask}, the difficulty of polyphone disambiguation in logographic languages mainly lies in heteronyms. Thus, the previous works~\cite{dai2019disambiguation,sun2019knowledge,park2020g2pm} usually use ``polyphone disambiguation'' to refer to ``heteronym disambiguation'' for those languages. In this work, we use ``polyphone disambiguation'' to refer to ``heteronym disambiguation'' following the previous works.

\section{The Detailed differences between NLR and Dict-TTS}
\label{diff_nlr}
The main differences between NLR and Dict-TTS can be summarized in the following aspected: 1) Attention design: Dict-TTS performs attention between each character and each pronunciation entry of the character separately, and explicitly resolves the polyphone, while NLR concatenates the dictionary texts of all the pronunciations of each character and attends jointly; 2) Dict-TTS further discusses the importance of aligning input representations in the semantic spaces; 3) Dict-TTS additionally leverages ASR-generated synthetic data; 4) Although both methods consider the goal of polyphone disambiguation, NLR is more focused on low-resource scenarios; 5) Dict-TTS is applied on PortaSpeech, while NLR on autoregressive transformer TTS. These improvements of Dict-TTS result in superior performance over NLR in the empirical studies.

\section{Analyses on Prosodic Realization}
\label{analyses_dur_pitch}
The DTW measure would be impacted by both lexical tone and prosodic realization. In order to further analyse the naturalness of prosody and rhythm for Dict-TTS and baseline systems, we evaluate the duration errors and character-level average pitch errors. For duration errors, we calculate the MSE of character-level durations. For character-level average pitch errors, we firstly calculate the mean pitch for each character's region in the mel spectrogram according to the Montreal Forced Aligner (MFA) to remove the influence of lexical tone, and then we calculate the MSE of the mean pitch sequences. The results on the Biaobei dataset are
shown in Table~\ref{table_dur_pitch}. It can be seen that the duration error and the average pitch error of Dict-TTS are significantly lower than those of the baseline systems, demonstrating the effectiveness of the extracted semantics from prior dictionary knowledge.

\begin{table}[hp]
  \caption{Duration error and average pitch error comparisons on the Biaobei dataset.}
  \label{table_dur_pitch}
  \centering
  \begin{tabular}{lll}
    \toprule
    Method           & Duration Error (ms)  & Average Pitch Error \\
    \midrule
    Character        &       36.2        &      1424.6 \\
    BERT Embedding   &       35.7        &      1312.1 \\
    NLR              &       36.4        &      1414.3 \\
    Phoneme (G2PM)   &       35.8        &      1341.7 \\
    Phoneme (pypinyin) &     35.3        &      1308.8 \\
    Dict-TTS         &  \textbf{34.4}    & \textbf{1232.3}  \\
    \bottomrule
  \end{tabular}
\end{table}

\section{Adding Rules to the Pronunciation Weights}
\label{adding_rules}
There are some pronunciation rules (like ``sandhi rules'') that can not be learned from the dictionary. For example, in Mandarin, \begin{CJK}{UTF8}{gbsn}"一"\end{CJK} before tone4 should be ``Y I2'' (e.g., \begin{CJK}{UTF8}{gbsn}``一段''\end{CJK}) and when \begin{CJK}{UTF8}{gbsn}``一''\end{CJK} is an ordinal word, it should be ``Y I1'' (e.g., \begin{CJK}{UTF8}{gbsn}``一四九五年''\end{CJK}). According to these pronunciation rules, we can obtain the correct pronunciation labels for some specific characters based on the part-of-speech (POS) tags of the input character sequence. After we obtain the correct pronunciation labels for these specific characters, we can directly force the pronunciation weights of these characters to be the ground truth values.

\section{Polyphone Disambiguation for Various Languages}
\label{variety_languages}
The polyphone disambiguation problem is critical in logographic languages such as Chinese, but is less problematic in phonograms like English.

For logographic languages like Chinese, although the lexicon can cover nearly all the characters, there are a lot of polyphones that can only be decided according to the context of a character. Thus, G2P conversion in this kind of languages is mainly responsible for polyphone disambiguation, which decides the appropriate pronunciation based on the current word context. Therefore, polyphone disambiguation is crucial in these languages and our method is an effective solution for the polyphone disambiguation problem in these languages. For alphabetic languages like English, lexicon cannot cover the pronunciations of all the words. Thus, the G2P conversion for English is mainly responsible for generating the pronunciations of out-of-vocabulary words~\cite{tan2021survey}. Although the polyphone disambiguation is less problematic in these languages, our methods can still be used as the modules to retrieve the correct pronunciation for polyphones and heteronyms in their G2P process (e.g., the Algorithm step 2 in \url{https://github.com/Kyubyong/g2p}).

\begin{CJK}{UTF8}{min}
In our experiments, JSUT dataset is a mixture of phonograms and logograms, which is different from Biaobei and Common Voice (HK). Japanese writing system consists of two types of characters: the kanji (漢字) and the syllabic kana – hiragana (平仮名) and katakana (片仮名). In our analysis, 32.42\% of the characters in JSUT dataset are kanji. The pronunciations of a part of the kanji can not only be specified by the semantic information and should be specified by empirical pronunciation rules. For example, most kanji (漢字) can be pronounced multiple ways: on-yomi (音読み) and kun-yomi (訓読み). Although the compound kanji usually uses on-yomi and one kanji probably uses kunyomi, the different readings are largely just chosen empirically in practice. Our Dict-TTS has the potential to work only for the kanji whose pronunciation should be specified based on the semantic meaning. Due to the characteristics of Japanese writing systems, in Table 1, although Dict-TTS surpasses the character-based system, it does not show comparable performance with the open source G2P module in Japanese. But as shown in Section 3.4, our method is compatible with the predefined rules from language experts by directly adding specific rules to pronunciation weight. We are sure that the performance of our method can be further improved by introducing the pronunciation rules in Japanese (e.g., the rules in the rule-based G2P baseline "pyopenjtalk").
\end{CJK}

\section{Potential Negative Societal Impacts}
\label{Negative}
Dict-TTS improves the pronunciation accuracy and prosody of the
synthesized speech voice and lowers the requirements for G2P conversion, which may cause unemployment
for people with related occupations. Besides, the production of fake speeches may cause voice security issues. Further efforts in automatic speaker verification should be made to improve voice security.

\end{document}